\documentclass[nofootinbib,twocolumn,preprintnumbers,prl]{revtex4-1}
\pdfoutput=1
\usepackage{amsmath,amsthm,amssymb,multirow,psfrag}
\usepackage{epsfig}
\usepackage{color}

\graphicspath{{./Figures/}}

\begin{document}

\def\lsim{\mathrel{\rlap{\lower4pt\hbox{\hskip1pt$\sim$}}
  \raise1pt\hbox{$<$}}}
\def\gsim{\mathrel{\rlap{\lower4pt\hbox{\hskip1pt$\sim$}}
  \raise1pt\hbox{$>$}}}
\newcommand{\vev}[1]{ \left\langle {#1} \right\rangle }
\newcommand{\bra}[1]{ \langle {#1} | }
\newcommand{\ket}[1]{ | {#1} \rangle }
\newcommand{\ev}{ {\rm eV} }
\newcommand{\kev}{{\rm keV}}
\newcommand{\mev}{{\rm MeV}}
\newcommand{\tev}{{\rm TeV}}
\newcommand{\mpl}{$M_{Pl}$}
\newcommand{\mw}{$M_{W}$}
\newcommand{\Ft}{F_{T}}
\newcommand{\Zparity}{\mathbb{Z}_2}
\newcommand{\BLambda}{\boldsymbol{\lambda}}
\newcommand{\met}{\;\not\!\!\!{E}_T}
\newcommand{\beq}{\begin{equation}}
\newcommand{\eeq}{\end{equation}}
\newcommand{\bea}{\begin{eqnarray}}
\newcommand{\eea}{\end{eqnarray}}
\newcommand{\nn}{\nonumber}
\newcommand{\gev}{{\mathrm GeV}}
\newcommand{\hc}{\mathrm{h.c.}}
\newcommand{\eps}{\epsilon}
\newcommand{\bwt}{\begin{widetext}}
\newcommand{\ewt}{\end{widetext}}
\newcommand{\draftnote}[1]{{\bf\color{blue} #1}}

\newcommand{\cO}{{\cal O}}
\newcommand{\cL}{{\cal L}}
\newcommand{\cM}{{\cal M}}

\newcommand{\fref}[1]{Fig.~\ref{fig:#1}} 
\newcommand{\eref}[1]{Eq.~\eqref{eq:#1}} 
\newcommand{\aref}[1]{Appendix~\ref{app:#1}}
\newcommand{\sref}[1]{Section~\ref{sec:#1}}
\newcommand{\tref}[1]{Table~\ref{tab:#1}}

\title{\LARGE{{\bf Golden Probe of the Top Yukawa}}}
\author{{\bf {Yi Chen$\,^{a}$,~Daniel Stolarski$\,^{b}$,~and Roberto Vega-Morales$\,^{c}$}}}

\affiliation{
$^a$Lauritsen Laboratory for High Energy Physics, California Institute of Technology, Pasadena, CA, USA\\
$^b$Theory Division, Physics Department, CERN, CH-1211 Geneva 23, Switzerland\\
$^c$Laboratoire de Physique Th\'{e}orique, CNRS - UMR 8627, Universit\'{e} Paris-Sud, Orsay, France
}

\email{
yichen@caltech.edu\\
daniel.stolarski@cern.ch\\
roberto.vega@th.u-psud.fr\\}

\begin{abstract}
We perform a preliminary study of the ability of the Higgs decay to four leptons to shed light on the top quark Yukawa couplings.~In particular we examine whether the $h\to 4\ell$ `golden channel' is sensitive to the $CP$ properties of the top quark couplings to the Higgs boson.~We show that kinematic distributions are sensitive to interference of the next-to-leading order electroweak corrections with the tree level $ZZ$ contribution.~This translates into a sensitivity to the top quark Yukawa couplings such that meaningful constraints on their $CP$ properties can begin to be obtained once $\sim 300$ fb$^{-1}$ of data has been collected at $\sim 14$ TeV, with significant improvements at higher luminosity or with a higher energy hadron collider.~This makes the $h\to4\ell$ channel a useful probe of the top quark Yukawa couplings that is qualitatively different from already established searches in $h\to V\gamma$ two body decays, $tth$, and $gg\to h$.~We also briefly discuss other potential possibilities for probing the top Yukawa $CP$ properties in $h\to2\ell\gamma$ and $\ell^+\ell^-\to h Z, h\gamma$.

\end{abstract}

\preprint{CERN-PH-TH-2015-106}

\maketitle

\section{Introduction} 
\label{sec:intro} 

The observation of a Higgs-like resonance with mass near 125 GeV~\cite{:2012gk,:2012gu} completes the Standard Model (SM) and opens up a vast new research program in studying its detailed properties in order to determine whether it is in fact the SM Higgs.~Direct study of the boson itself is the best way to unravel the nature of this new state and answer interesting questions such as whether its interactions violate $CP$.~It has been established that its couplings to $ZZ$ are dominantly $CP$ even~\cite{Aad:2013xqa,Chatrchyan:2013mxa}, but $CP$ is violated in nature, so if there is physics beyond the SM (BSM), some Higgs couplings may not conserve $CP$.

In the SM, the largest coupling of the Higgs is to the top quark.~Therefore, studying the Higgs top system is particularly interesting because it could be an ideal place to discover new physics.~Furthermore, because of the size of this coupling, the hierarchy problem is sharpest in the top sector, so potential solutions to the hierarchy problem could easily modify the couplings between the Higgs and the top.~This coupling can be studied directly using the $tth$ production rate, which as yet is unobserved~\cite{Khachatryan:2014qaa,ATLAS-CONF-2014-011,Aad:2014lma}.~Various studies have also shown that kinematic observables can be constructed to study the size and $CP$ properties of the top Yukawa in this channel at the LHC~\cite{Gunion:1996xu,Gunion:1998hm,Ellis:2013yxa,Khatibi:2014bsa,He:2014xla,Boudjema:2015nda}, though it requires measurements of top and Higgs decays which may be difficult in the high luminosity environment of the LHC.

The Higgs decay to photons is mediated by a top quark loop (and the larger $W$ loop), so this channel can also probe the top Yukawa coupling.~Similarly, the cross section of Higgs production via the gluon fusion process is sensitive to the top Yukawa coupling.~Therefore global fits using rates can be used to constrain it under various assumptions.~This has been done by the experimental collaborations~\cite{Khachatryan:2014jba,ATLAS-CONF-2014-010}, as well as by several theoretical groups~\cite{Carmi:2012in,Banerjee:2012xc,Plehn:2012iz,Djouadi:2012rh,Belanger:2012gc,Cheung:2013kla,Falkowski:2013dza,Giardino:2013bma,Ellis:2013lra,Bernon:2014vta} which indicate a top Yukawa coupling consistent with the SM.~Even with the various assumptions in these analyses, sizable deviations from the SM prediction are still allowed.

Additional probes of the $CP$ properties of the top Higgs system include production of a Higgs in association with a single top~\cite{Tait:2000sh,Ellis:2013yxa,Agashe:2013hma,Chang:2014rfa,Yue:2014tya,Demartin:2015uha}, kinematic distributions in gluon fusion Higgs production~\cite{Grojean:2013nya,Demartin:2014fia}, and low energy $CP$ violating observable such as EDMs~\cite{Brod:2013cka}.~All of these probes, however, are indirect and suffer from significant inverse problems.~Namely, if deviations from SM predictions are discovered, it is very difficult to determine if they are coming from modifications to the top Yukawa coupling, or some other type of new physics.~Therefore, it is important to have as many complementary probes as possible. 

In this work we propose a new avenue to study the top-Higgs system:~the Higgs decay to four leptons.~This so-called `golden channel' has already been used extensively to study the spin of the Higgs as well as the $CP$ and tensor structure of its coupling to gauge boson pairs~\cite{Nelson:1986ki,Soni:1993jc,Chang:1993jy,Barger:1993wt,Arens:1994wd,Choi:2002jk,Buszello:2002uu,Godbole:2007cn,Kovalchuk:2008zz,Cao:2009ah,Gao:2010qx,DeRujula:2010ys,Gainer:2011xz,Campbell:2012cz,Campbell:2012ct,Belyaev:2012qa,Coleppa:2012eh,Bolognesi:2012mm,Boughezal:2012tz,Stolarski:2012ps,Avery:2012um,Chen:2012jy,Modak:2013sb,Gainer:2013rxa,Grinstein:2013vsa,Sun:2013yra,Anderson:2013fba,Chen:2013waa,Buchalla:2013mpa,Chen:2013ejz,Gainer:2014hha,Chen:2014gka,Chen:2014pia,Chen:2015iha,Bhattacherjee:2015xra,Gonzalez-Alonso:2015bha}.~The leading contribution to the golden channel comes from the tree level coupling of the Higgs to $ZZ$ generated during electroweak symmetry breaking (EWSB).~At one loop, however, additional couplings of the Higgs to $Z\gamma$ and $\gamma\gamma$ pairs (as well as $ZZ$) can be generated and mediate Higgs decays to four leptons.

In the SM, these next-to-leading order (NLO) contributions are dominated by $W$ and top loops.~While these are one-loop contributions, the large available phase space for the $Z\gamma$ and $\gamma\gamma$ intermediate states as well as the differential spectra allow for these one loop contributions to be distinguished from the tree level $ZZ$ coupling~\cite{Stolarski:2012ps,Chen:2012jy,Chen:2013ejz,Chen:2014gka,Chen:2015iha}.~In particular, due to interference effects between the higher dimensional $Z\gamma$ and $\gamma\gamma$ couplings with the tree level $ZZ$ coupling, the $h\to4\ell$ ($4\ell \equiv 2e2\mu, 4e, 4\mu$) channel is surprisingly sensitive to the $CP$ properties of these loop induced couplings, especially for $\gamma\gamma$~\cite{Chen:2014gka,Chen:2015iha}.

In this work, we exploit the fact that the top quark mediates Higgs decays to both $\gamma\gamma$ and $Z\gamma$ intermediate states via the same couplings to the Higgs boson.~Therefore, unlike previous work which focused on measuring higher dimension effective couplings to gauge bosons, we here use the underlying loop processes to gain sensitivity to the physical parameters of the SM or BSM effects.~Furthermore, because the one-loop top mediated effects interfere with the tree-level diagram, the differential cross section has a component which is linearly sensitive to $CP$ violation and only contains one power of the loop factor.~This is in contrast to $h\to\gamma\gamma$ and $h\to Z\gamma$ two body decays or $tth$ rate measurements which are sensitive only to the sum of squares of the $CP$ even and odd components of the top Yukawa coupling.~Therefore, if the coupling of the Higgs to the top has the wrong sign, as can happen in certain two Higgs doublet~\cite{Ginzburg:2001ss,Dumont:2014wha} and triplet~\cite{Hedri:2013wea} models, or if there is a non-trivial $CP$ phase, this can in principle be observed directly using \emph{only} the golden channel independently of these other measurements. 

Here we perform an initial feasibility study to explore whether the golden channel can be used as a probe of the Higgs top quark Yukawa coupling and perhaps uncover $CP$ violation.~To do this we utilize the parameter extraction framework developed in~\cite{Chen:2012jy,Chen:2013ejz,Chen:2014pia,Chen:2014gka,Chen:2014hqs,Chen:2015iha} to study effective Higgs couplings and adapt it to include the leading contributions from top quark (and $W$) loop effects.~We demonstrate a proof of principle that the $h\to4\ell$ channel has the potential to probe the $CP$ properties of the top Yukawa at the LHC with very promising prospects at a future higher energy hadron collider.~We also briefly discuss other potential possibilities for probing the top Yukawa in $h\to2\ell\gamma$ and $\ell^+\ell^-\to h Z, h\gamma$.

\section{Probing the Top Yukawa in $h\to4\ell$}
\label{sec:yukawa}

Many previous studies of the golden channel have focused on probing effective couplings of the Higgs to gauge bosons of the form,
\bea
\mathcal{L}_{VV'} 
&\sim& \frac{h}{v}
\Big(
 A_{1}^{ZZ} m_Z^2 \, Z^{\mu}Z_{\nu}
 + A_{2}^{VV'} V^{\mu\nu}V'_{\mu\nu} \nn\\
&+& A_{3}^{VV'} V^{\mu\nu} \widetilde{V'}_{\mu\nu}
+ A_{4}^{ZV} \partial^\mu Z^\nu V_{\mu\nu}  \Big) \, ,
\label{eq:eff-op}
\eea
where $V,V' = Z, \gamma$, and $V^{\mu\nu}$ ($\tilde{V}^{\mu\nu}$) is the usual field strength (dual field strength) tensor.~These mediate Higgs decays to four leptons via the diagram shown in~\fref{hto4l}.~The differential distributions for the many kinematic observables in $h\to4\ell$~\cite{Chen:2012jy,Chen:2013ejz,Chen:2014pia,Chen:2014gka} give us a probe into detailed properties of these effective couplings.~In particular, it was demonstrated in~\cite{Chen:2014gka,Chen:2015iha} that golden channel measurements are surprisingly sensitive to the effective couplings of the Higgs boson to $Z\gamma$ and $\gamma\gamma$ pairs.~Because of shape differences and interference with the tree level $ZZ$ coupling, the sensitivity is strong enough that SM values of the $\gamma\gamma$ effective couplings should be probed well before the end of LHC running.~Prospects for $Z\gamma$ are less promising, but still perhaps possible at the LHC and very promising at a future $100$~TeV collider.~This motivates the question of whether the sensitivity to these effective couplings translates into sensitivity to the underlying loop processes.
\begin{figure}[tb]
\centering
\begin{minipage}[c]{0.3\textwidth}
\includegraphics[width=1.1\textwidth]{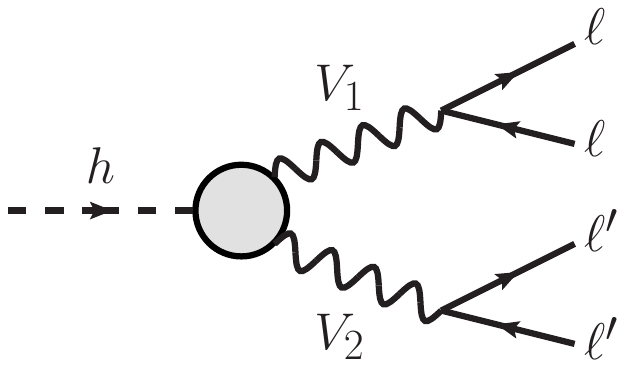}
\end{minipage}
\hfill
\caption{Schematic representation of the $hVV$ corrections to the $h\to4\ell$ amplitude where $V_{1,2} = Z,\gamma$ and $\ell = e, \mu$.}
\label{fig:hto4l}
\end{figure}

In the SM, the $A_2^{V\gamma}$ couplings are generated at one loop dominantly through a $W$ boson loop followed by the smaller top loop contribution shown in \fref{Feyn}, while $A_3^{V\gamma}$ is zero at this order.~The $A_4^{Z\gamma}$ coupling is generated at one loop, but vanishes for an on-shell photon and to leading order in the heavy loop particle expansion.~The leading $W$-loop contribution to $A_2^{V\gamma}$ involves parameters such as the $W$ mass and gauge couplings that are well measured from LEP~\cite{ALEPH:2005ab,Schael:2013ita} and the LHC~\cite{:2012gk,:2012gu,Chatrchyan:2013fya,Aad:2014mda} experiments.~Therefore, it is a reasonable approximation to take these $W$ loops to be fixed during our parameter extraction of the top Yukawa coupling.~Studying the sensitivity in $h\to4\ell$ to electroweak parameters in the $W$ loops would also be interesting, but requires a more careful treatment of other SM one-loop contributions and so is left to ongoing work~\cite{followup2}.
\begin{figure*}[tb]
\centering
\begin{minipage}[c]{0.3\textwidth}
\includegraphics[width=0.97\textwidth]{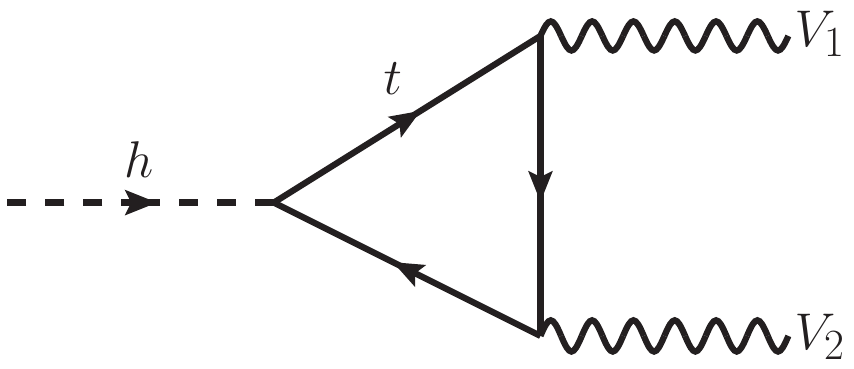}
\end{minipage}
\begin{minipage}[c]{0.3\textwidth}
\includegraphics[width=0.97\textwidth]{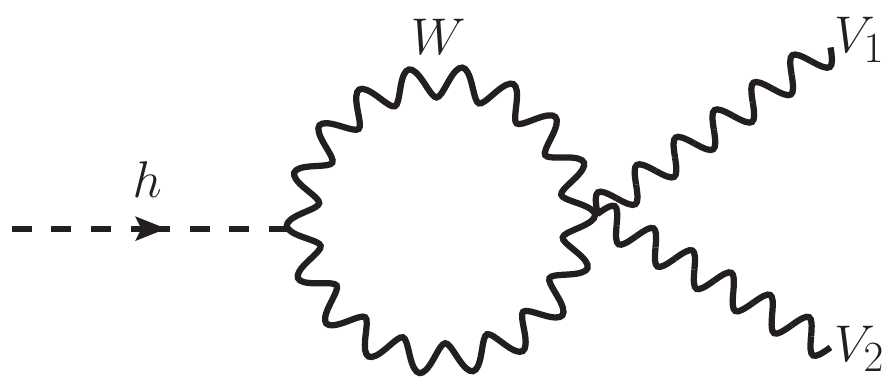}
\end{minipage}
\begin{minipage}[c]{0.3\textwidth}
\includegraphics[width=0.97\textwidth]{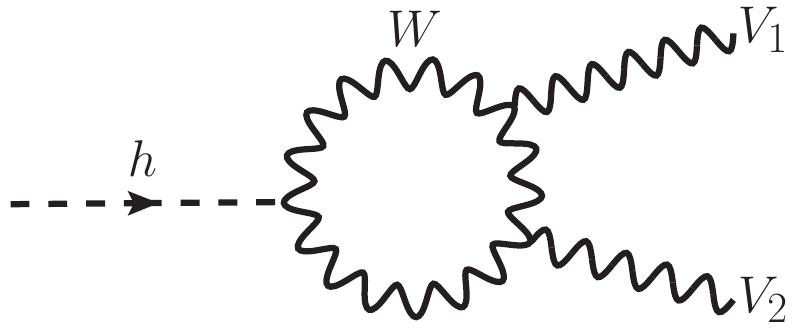}
\end{minipage}
\hfill
\caption{One-loop contributions from top quark (left) and $W$ boson to $h \to V_1 V_2 \to 4\ell~(V_i = Z, \gamma)$.}
\label{fig:Feyn}
\end{figure*}

The top loop on the other hand involves various parameters which are not as precisely constrained.~Furthermore, axial couplings between the Higgs and the top quark can generate $A_3^{Z\gamma}$ effective couplings which are vanishingly small in the SM.~Focusing on the top Yukawa, we take the top pole mass and the $ttZ$ coupling to be fixed, though it would be interesting to study these as well.~We parametrize the top Yukawa couplings as,
\bea
\label{eq:Ylag}
\mathcal{L}_t \supset \frac{m_t}{v} h \bar{t} ( y_t  + i \tilde{y}_t \gamma^5 )t ,
\eea
where $m_t$ is defined to be the \emph{pole mass} found in the top quark propagator with $y_t=1,\tilde{y}_t = 0$ at tree level in the SM.~Note that the large pole mass of the top leads to little sensitivity to the top mass in practice in $h\to VV$ decays.~This is equivalent to saying the top contribution is well approximated by a constant effective $hVV$ coupling after the top has been integrated out.~Thus whether we fix or allow the top pole mass to vary makes a negligible difference on our results.

After the $W$ and top, the next largest contribution to the effective $Z\gamma$ and $\gamma\gamma$ couplings comes from the bottom quark contribution.~This effect is suppressed by $\sim (m_b/m_t)^2$ in the matrix element relative to the top contribution which is itself subdominant to the $W$ loop.~Thus, to a very good approximation, the $Z\gamma$ and $\gamma\gamma$ effective couplings \emph{only} receive contributions at one-loop from the $W$ boson and top quark.

The $h\rightarrow 4\ell$ process receives additional one-loop electroweak (EW) corrections that are not of the form shown in~\fref{hto4l}.~Since the $Z\gamma$ and $\gamma\gamma$ effective couplings in~\eref{eff-op} are only first generated at one loop, they do not receive a contribution from these additional EW corrections at this loop order.~These include processes such as corrections to the $Z$ propagator and coupling to leptons as well as various other non-local interactions all of which are computable~\cite{Bredenstein:2006rh,Bredenstein:2006nk}.~Thus in principle we can make a precise prediction for all contributions not involving the top Yukawa coupling.~This allows us to treat this part of the amplitude which does not depend on the top Yukawa as part of the SM `background' to our top Yukawa `signal'.

\subsection{Discussion of Signal and `Backgrounds'}

To be more explicit, we can write the $h\to4\ell$ amplitude up to one loop as follows,
\bea
\mathcal{M}_{4\ell} = \mathcal{M}_{SM}^0 + \mathcal{M}_{EW}^1 + \mathcal{M}_{t}^1   \, .
\eea
The leading term $\mathcal{M}_{SM}^0 $ arises from the tree level $hZZ$ coupling,
\bea
\label{eq:LSM0}
\mathcal{L}_{SM}^0 \supset \frac{m_Z^2}{v} h Z^\mu Z_\mu ,
\eea
which is generated during EWSB and is responsible for giving the $Z$ boson its mass.~The second term $\mathcal{M}_{EW}^1$ involves all SM one-loop contributions \emph{independent} of the top Yukawa, though there are one-loop corrections from top quark loops to the $Z$ boson propagator for example.~Finally, $\mathcal{M}_{t}^1$ encodes the one-loop contribution sensitive to the top Yukawa coupling and which enters via the first diagram in~\fref{Feyn}.~In this work, we will treat $\mathcal{M}_{t}^1$ as our signal and fit for the parameters in \eref{Ylag}, while we will treat the rest of the matrix element as `background' which we keep fixed.~There are also real non-Higgs backgrounds, whose leading contributions must be accounted for as well and will be discussed below.

We can further characterize the `background' in $\mathcal{M}_{EW}^1$ by isolating those contributions which are generated by $hVV$ (where $VV = ZZ, Z\gamma, \gamma\gamma$) effective couplings of the form shown in \fref{hto4l} to write,
\bea
\mathcal{M}_{EW}^1 = \bar{\mathcal{M}}_{EW}^1 +  \mathcal{M}_{EW}^{VV},
\eea
where we have defined,
\bea
\mathcal{M}_{EW}^{VV} = \mathcal{M}_{EW}^{ZZ} + \mathcal{M}_{EW}^{Z\gamma} + \mathcal{M}_{EW}^{\gamma\gamma}.
\eea
These contributions all have the form of~\fref{hto4l} and will be examined more closely below.

There are many contributions to $\bar{\mathcal{M}}_{EW}^1 $, all of which are computable and can in principle be extracted from~\cite{Bredenstein:2006rh,Bredenstein:2006nk}.~Some of these one loop contributions can be absorbed into shifts of the tree level couplings.~Others can be modeled using effective operators.~There are also real photon emission effects in $h\to 4\ell$~\cite{Bredenstein:2006rh,Bredenstein:2006nk,Boselli:2015aha} which can be non-negligible in certain regions of phase space, but which can also be included~\cite{Gonzalez-Alonso:2014eva}.~The key point however is that these corrections do not depend on the top Yukawa, allowing us to treat them as fixed when fitting for the top Yukawa.~Furthermore, since at one loop these corrections do not contribute to the $Z\gamma$ or $\gamma\gamma$ effective couplings to which we are most sensitive in $h\to4\ell$~\cite{Chen:2014gka,Chen:2015iha}, and since they are sub-dominant over most of the phase space~\cite{Gonzalez-Alonso:2014eva}, we will neglect them in this preliminary study.~However, a detailed investigation of their effects is worthwhile and will be done in future work.~Thus in the end, for the present study we define the Higgs part of our `background' (in contrast to non-Higgs background to be discussed) as,
\bea
\label{eq:LBG}
\mathcal{M}^h_{BG} = \mathcal{M}^{0}_{SM}  + \mathcal{M}_{EW}^{VV} .
\eea
This part of the $h\to4\ell$ amplitude will be treated as fixed during the parameter extraction procedure.

As mentioned, our `signal' is then the top quark loop in the $Z\gamma$ and $\gamma\gamma$ effective couplings which we call $\mathcal{M}_{t}^{Z\gamma}$ and $\mathcal{M}_{t}^{\gamma\gamma}$.~Of course a top quark loop will also generate $\mathcal{M}_{t}^{ZZ}$ via $hZZ$ effective couplings, though in practice the sensitivity to this contribution is an order of magnitude weaker than for the $Z\gamma$ and $\gamma\gamma$ effective couplings~\cite{Chen:2014gka,Chen:2015iha}.~Thus our final signal involving a top loop can be written as,
\bea
\label{eq:LS}
\mathcal{M}_{t}^{1} = \mathcal{M}_{t}^{ZZ} +  \mathcal{M}_{t}^{Z\gamma} + \mathcal{M}_{t}^{\gamma\gamma} .
\eea
All of the contributions in~\eref{LBG} and~\eref{LS} enter the $h\to4\ell$ amplitude via the $hVV$ couplings and can be represented by the diagram in~\fref{hto4l}.~Thus, by focusing on these and neglecting $\bar{\mathcal{M}}_{SM}^{1}$ we are assuming in the present study that any deviations from the tree level SM prediction occur only through loops which generate the $hVV$ effective couplings.~As discussed above this is a reasonable approximation for current purposes.~We will examine the contributions in~\eref{LBG} and~\eref{LS} more closely below.

There is of course a non-Higgs background which comes dominantly from the continuum $q\bar{q}\to4\ell$ process~\cite{Khachatryan:2014kca} and can have important effects.~As discussed in~\cite{Chen:2015iha} this background enters almost entirely due to detector resolution effects.~If detectors had perfect energy resolution, the signal region would essentially be a $\delta$-function centered at the Higgs mass leading to an effectively background free sample.~However, imperfect detector resolution has the effect of widening the signal region, thus introducing more non-Higgs background into the sample and degrading the sensitivity to the $hVV$ effective couplings~\cite{Chen:2015iha}.

For this $q\bar{q}\to4\ell$ background we utilize the analytic expressions computed in~\cite{Chen:2012jy,Chen:2013ejz} and follow the procedure in~\cite{Gainer:2011xz,Chen:2015iha} to build a signal plus background likelihood which includes the parton distribution functions as well as crude modeling of detector resolution effects.~More details on this implementation can be found in~\cite{Gainer:2011xz,Chen:2012jy,Chen:2013ejz,Chen:2015iha}.~For a more realistic analysis, careful treatment of detector resolution and additional background effects can be done with the framework in~\cite{Chen:2014pia,Chen:2014hqs,Khachatryan:2014kca}, but is left to future work.~However, these detector effects are not expected to qualitatively change the results obtained here.

\subsection{The top and $W$ Loops}

Restricting our attention to the loops which generate the $hVV$ effective couplings in~\eref{LBG} and~\eref{LS}, the matrix element for the $h\rightarrow 4\ell$ decay can be written as,
\bea
\mathcal{M}(h\rightarrow 4\ell) &=& \mathcal{M}^{\mu\nu}(h\rightarrow V_1 V_2) \times \nonumber\\
\mathcal{P}_{\mu\alpha}(V_1)&\mathcal{M}^{\alpha}&(V_1\rightarrow 2\ell)\mathcal{P}_{\nu\beta}(V_2)\mathcal{M}^{\beta}(V_2\rightarrow 2\ell)\, ,
\label{eq:matrixelement}
\eea
where $V=Z,\gamma$, and $\mathcal{P}_{\mu\nu}(V_i)$ are the propagators of the vector bosons.~The second line is described by well measured physics of vector bosons coupling to leptons, while the matrix element on the first line encodes all the Higgs physics and for which constraints are far weaker. 

The $h\to V_1 V_2$ matrix element can be decomposed into the following tensor structure,
\bea
&&\mathcal{M}^{\mu\nu}(h\rightarrow V_1 V_2) =
\label{eq:MhVV}
\frac{1}{v}
\mathcal{C}_1^{i} m_Z^2 g^{\mu\nu} + \\
&&~~~~\frac{1}{v} \mathcal{C}_2^{i} (k_1^\nu k_2^\mu - k_1\cdot k_2 g^{\mu\nu})
+ \frac{1}{v}\mathcal{C}_3^{i} \epsilon^{\mu\nu\alpha\beta} k_{1\alpha} k_{2\beta}
,\nn
\eea
where $i = ZZ, Z\gamma, \gamma\gamma$ and $k_1$ and $k_2$ represent the four momenta of the intermediate vector bosons (or lepton pairs).~The Lorentz invariant form factors $\mathcal{C}_n^i$ are in general momentum dependent for the off-shell intermediate vector bosons and have the generic form,
\bea
\label{eq:Atof}
\mathcal{C}_n^i \sim g_X f_i(m_h^2/m_X^2, k_1^2, k_2^2) ,
\eea
where $f_i(m_h^2/m_X^2, k_1^2, k_2^2)$ is the loop function for underlying particle $X$ with coupling to the Higgs $g_X$.~For a $\sim 125$~GeV Higgs mass, the dependance on $k_i^2$ is rather weak~\cite{Gonzalez-Alonso:2014eva} over much of the phase space and, to a sufficiently good approximation, the $\mathcal{C}_n^i$ are given by setting $k_i^2$ equal to the physical mass of the relevant gauge boson.~The $k_i^2$ dependence of the form factors can be relevant in certain regions of phase space and factoring it in may aid in sensitivity, warranting closer examination.~However, in this initial study we seek to first establish a proof of principle with the leading terms leaving a more detailed exploration of these `off-shell' effects to currently ongoing work~\cite{followup2}.

Thus, the form factors $\mathcal{C}_{2,3}^{V\gamma}$ in~\eref{MhVV} will be the ones that control Higgs decay to \textit{on-shell} $\gamma\gamma$ and $Z\gamma$ pairs.~The leading contributions to these form factors comes from $W$ and top loops which are shown in Fig~\ref{fig:Feyn}.~These one-loop contributions have been computed for $h\to Z\gamma$~\cite{Cahn:1978nz,Bergstrom:1985hp} and $h\to\gamma\gamma$~\cite{Ellis:1975ap,Shifman:1979eb} (including pseudoscalar couplings~\cite{Weiler:1988xn} for the top) and can be straightforwardly incorporated into the analytic expressions for the $h\to4\ell$ fully differential cross section computed in~\cite{Chen:2012jy,Chen:2013ejz}.~For our explicit expressions of the top and $W$ loop functions, we use the conventions in~\cite{Djouadi:2005gi,Djouadi:2005gj}.

As discussed above, the sensitivity to the higher dimensional $hZZ$ effective couplings is significantly weaker than for the $hZ\gamma$ and $h\gamma\gamma$ effective couplings~\cite{Chen:2015iha}.~Furthermore, though the $hZZ$ effective couplings receive contributions from top and $W$ loops, there are also a number of other one-loop contributions involving $Z$ and Higgs bosons.~The already weak sensitivity to these $hZZ$ couplings makes disentangling the top contribution from other contributions difficult.~We therefore simply will model these with the set of dimension 5 operators:
\bea
\label{eq:ZZlag}
\mathcal{L}_{ZZ} &\supset& \frac{h}{4v}
\Big(
 A_{2}^{ZZ} Z^{\mu\nu}Z_{\mu\nu}
+ A_{3}^{ZZ} Z^{\mu\nu} \widetilde{Z}_{\mu\nu} \nn\\
&&~~~~+ 4 A_{4}^{ZZ} \partial_\mu Z_\nu Z^{\mu\nu} 
\Big) \, ,
\eea
where the $A_n^{ZZ}$ are taken as real and constant.~To study the potential effects of these contributions we treat $A_n^{ZZ}$ as nuisance parameters in our parameter extraction procedure allowing them to vary along with the top quark Yukawa.~As we will see, the effects of the operators in~\eref{ZZlag} do not greatly affect our sensitivity to the top Yukawa, especially once sufficient statistics are accumulated.

\subsection{Other Possible Probes of the Top Yukawa}

In~\cite{Chen:2014ona} it was shown that due to weak phase/strong phase interference effects, the three body $h\to2\ell\gamma$ decay is also sensitive to the $CP$ violation in the effective $hZ\gamma$ and $h\gamma\gamma$ couplings.~Thus probing the $CP$ properties of the top Yukawa may also be possible in this channel at the LHC or future hadron collider.~Since this channel is less sensitive and requires an understanding of the much larger backgrounds than in $h\to4\ell$, we do not examine this possibility in detail here.

Crossing symmetry implies $\ell^+\ell^- \to h Z, h\gamma$ scattering at a future lepton collider~\cite{Chen:2014ona,Shen:2015pha} may also be capable of probing the top Yukawa $CP$ properties.~Recently it has also been shown that interference between signal and background can be used to probe the effective $hZ\gamma$ and $h\gamma\gamma$ couplings in $gg\to 2\ell\gamma$~\cite{Farina:2015dua}, which implies this may also be used to probe the top Yukawa.~We leave an investigation of these interesting possibilities to future work.

\section{Sensitivity at LHC and Beyond}
\label{sec:lhc}

We now quantitatively explore the feasibility of the LHC or a future hadron collider to probe the $CP$ properties of the top Yukawa coupling in $h\to4\ell$.~In particular, we estimate approximately how many events will be needed in $h\to4\ell$ to begin probing values of Yukawa couplings which are of the same order as the $\mathcal{O}(1)$ SM prediction.~We also examine approximately at what point $h\to4\ell$ will become relevant as a measurement relative to $h\to V\gamma$ and $tth$ searches for studying the top Yukawa (we will not consider $gg\to h$, but see~\cite{Carmi:2012in,Banerjee:2012xc,Plehn:2012iz,Djouadi:2012rh,Belanger:2012gc,Cheung:2013kla,Falkowski:2013dza,Giardino:2013bma,Ellis:2013lra,Bernon:2014vta} for various studies of this channel).~Once this level of sensitivity is reached, a more complete analysis including the various other one-loop corrections discussed above will need to be conducted in order to give precise constraints on the top Yukawa.

For all results in the present study we have utilized the Higgs effective couplings extraction framework developed in~\cite{Chen:2012jy,Chen:2013ejz,Chen:2014pia,Chen:2014gka} which incorporates all observables available in the (normalized) $h\to4\ell$ fully differential decay width and adapted it to include the top and $W$ loop functions discussed above.~Also as discussed, we include the dominant $q\bar{q}\to4\ell$ background and a crude modeling of detector resolution~\cite{Chen:2015iha}.~For the Higgs signal, this includes a smearing of the four lepton invariant mass ($M_{4\ell}$) distribution with a gaussian of $\sigma = 2$~GeV centered at the Higgs mass which we take to be $125$~GeV.~Note that these resolution effects also enter into the lepton pair invariant masses ($M_{\ell\ell}$).~Following the procedure in~\cite{Gainer:2011xz}, the parton level differential cross sections for $h\to4\ell$ and $q\bar{q}\to4\ell$ are combined with the (CTEQ6l1~\cite{Lai:1999wy,Pumplin:2002vw}) parton distributions for the $gg$ and $q\bar{q}$ initial states.~Further details and validation of this procedure with MadGraph5\_aMC@NLO~\cite{Alwall:2014hca} can be found in~\cite{Chen:2013ejz,Chen:2014gka}.

\subsection{Parameter and Phase Space Definition}

Before presenting our results, we first define our parameter and phase space.~As discussed above, in order to study the effects of some of the one-loop contributions we have not computed which enter through the $ZZ$ sector, we allow the higher dimensional effective $ZZ$ couplings in~\eref{ZZlag} to vary in the fitting procedure.~Thus we define our multi-dimensional parameter space as,
\bea
\label{eq:params}
\vec{\lambda} = (y_t, \tilde{y}_t | A_{2}^{ZZ}, A_{3}^{ZZ}, A_{4}^{ZZ}) .
\eea
Note in particular that we are taking the tree level $hZZ$ coupling as fixed and equal to its SM value in~\eref{LSM0}.

To estimate the sensitivity we obtain what we call an `effective' $\sigma(\lambda)$ or \emph{average error} defined in~\cite{Chen:2015iha} as,
\begin{eqnarray}
\label{eq:sigma}
\sigma(\lambda) = \sqrt{\frac{\pi}{2}} \langle |\hat{\lambda} - \vec{\lambda}_o| \rangle ,
\end{eqnarray}
where $\hat{\lambda}$ is the value of the best fit parameter point obtained by maximization of the likelihood with respect to~$\vec{\lambda}$.~Here $\vec{\lambda}_o$ represents the `true' value with which our data sets are generated utilizing a MadGraph5\_aMC@NLO~\cite{Alwall:2014hca} implementation of the effective $hVV$ couplings~\cite{Chen:2012jy,Chen:2013ejz}.~The average error is then found by conducting a large number of pseudoexperiments with a fixed number of events and obtaining a distribution for $\hat{\lambda}$ which will have some spread centered around the average value.~We then translate the width of this distribution into our effective $\sigma(\lambda) $ which converges to the usual interpretation of $\sigma(\lambda) $ when the distribution for $\hat{\lambda}$ is perfectly gaussian.~We repeat this procedure for a range of number of signal events ($N_S$) to obtain $\sigma(\lambda) $ as a function of $N_S$.

Following the strategy proposed in~\cite{Chen:2015iha}, we will use a set of phase space cuts which are optimized for sensitivity to the $Z\gamma$ and $\gamma\gamma$ effective couplings.~These cuts were shown to greatly improve the sensitivity to the $Z\gamma$ and $\gamma\gamma$ effective couplings over currently used CMS cuts~\cite{CMS-PAS-HIG-14-014,Khachatryan:2014kca}.~They are defined as:
\begin{itemize}
	\item 115~GeV $< M_{4\ell} < 135$~GeV
	\item $p_T > (20,10,5,5)$~GeV for lepton $p_T$ ordering,
	\item $|\eta_\ell| < 2.4$ for the lepton rapidity,
	\item $M_{\ell\ell} > 4$~GeV, $M_{\ell\ell}\rm{(OSSF)} \notin (8.8, 10.8)$~GeV,
\end{itemize}
where $M_{\ell\ell}$ are all six lepton pair invariant masses and we explicitly remove events with opposite sign same flavor (OSSF) lepton pairs that have $M_{\ell\ell}$ in the range $8.8-10.8$~GeV in order to avoid contamination from $\Upsilon$ QCD resonances.~We refer to these as `Relaxed$-\Upsilon$' cuts.

While these cuts perform significantly better in terms of sensitivity to the effective $hZ\gamma$ and $h\gamma\gamma$ couplings than the currently used CMS cuts~\cite{Chen:2015iha}, they also allow more non-Higgs background into the sample.~It is therefore necessary to include the dominant non-Higgs $q\bar{q}\to 4\ell$ background discussed above as it can have a significant effect on parameter extraction when these cuts are utilized.~To do this we combine the background and signal into a single likelihood and fit for the background fraction during the parameter extraction procedure along with the parameters in~\eref{params}.~The background fractions used during event generation can be found in~\cite{Chen:2015iha}.~Many more details on the various aspects of the parameter extraction framework including the building of the signal plus background likelihood and the fitting procedure can be found in~\cite{Gainer:2011xz,Chen:2012jy,Chen:2013ejz,Chen:2014pia,Chen:2014gka}.

We also comment that for these cuts some of one-loop EW corrections we have neglected~\cite{Bredenstein:2006rh,Bredenstein:2006nk,Boselli:2015aha} may become relevant.~For this reason we also will discuss results utilizing CMS-like cuts~\cite{Chen:2015iha} for which these contributions are phase space suppressed~\cite{Gonzalez-Alonso:2014eva}, but this will not qualitatively affect the discussion.
%

\subsection{Sensitivity as Function of Luminosity}

In~\fref{curves} we show sensitivity curves for $\sigma(y_t)$ (red) and $\sigma(\tilde{y}_t)$ (blue) as function of the number of signal events ($N_S$) (bottom axis) and luminosity $\times$ efficiency (top axis) assuming SM production ($gg\to h$ plus VBF at 14 TeV) and branching ratios~\cite{Dittmaier:2011ti,Heinemeyer:2013tqa}.~In these fits we have utilized the Relaxed$-\Upsilon$ cuts discussed above and include both signal and the dominant $q\bar{q}\to 4\ell$ background.~We have combined the $2e2\mu, 4e, 4\mu$ channels and fit to a `true' point of $\vec{\lambda} = (1,0 | 0.01,0,0.007)$ corresponding to the SM prediction for the top Yukawa which is indicated by the dotted black line.

We see stronger sensitivity to the axial coupling $\tilde{y}_t$ than to the vector-like coupling $y_t$.~This is because the $CP$ even component of the top loop is dominated by the $W$ loop, but the $CP$ odd couplings $\tilde{y}_t$ does not have to compete with an analogous $W$ contribution.~We also study the effect of floating the effective $ZZ$ couplings (solid curves) defined in~\eref{ZZlag}, versus holding these couplings fixed (dashed curves).~The values chosen for these $ZZ$ effective couplings are only representative and whether we take their true value to be zero or $\mathcal{O}(10^{-2})$ makes negligible difference since the sensitivity to these couplings is weak~\cite{Chen:2014pia,Chen:2015iha}.~What is important to establish is whether allowing them to vary in the fit affects the sensitivity to the top Yukawa.~We see clearly in~\fref{curves} that this effect is small as expected from differences in the kinematic shapes of the $ZZ$, $Z\gamma$, and $\gamma\gamma$ intermediate states~\cite{Chen:2014pia,Chen:2015iha}.~
\begin{figure}[tbh]
\begin{center}
\includegraphics[width=.45\textwidth]{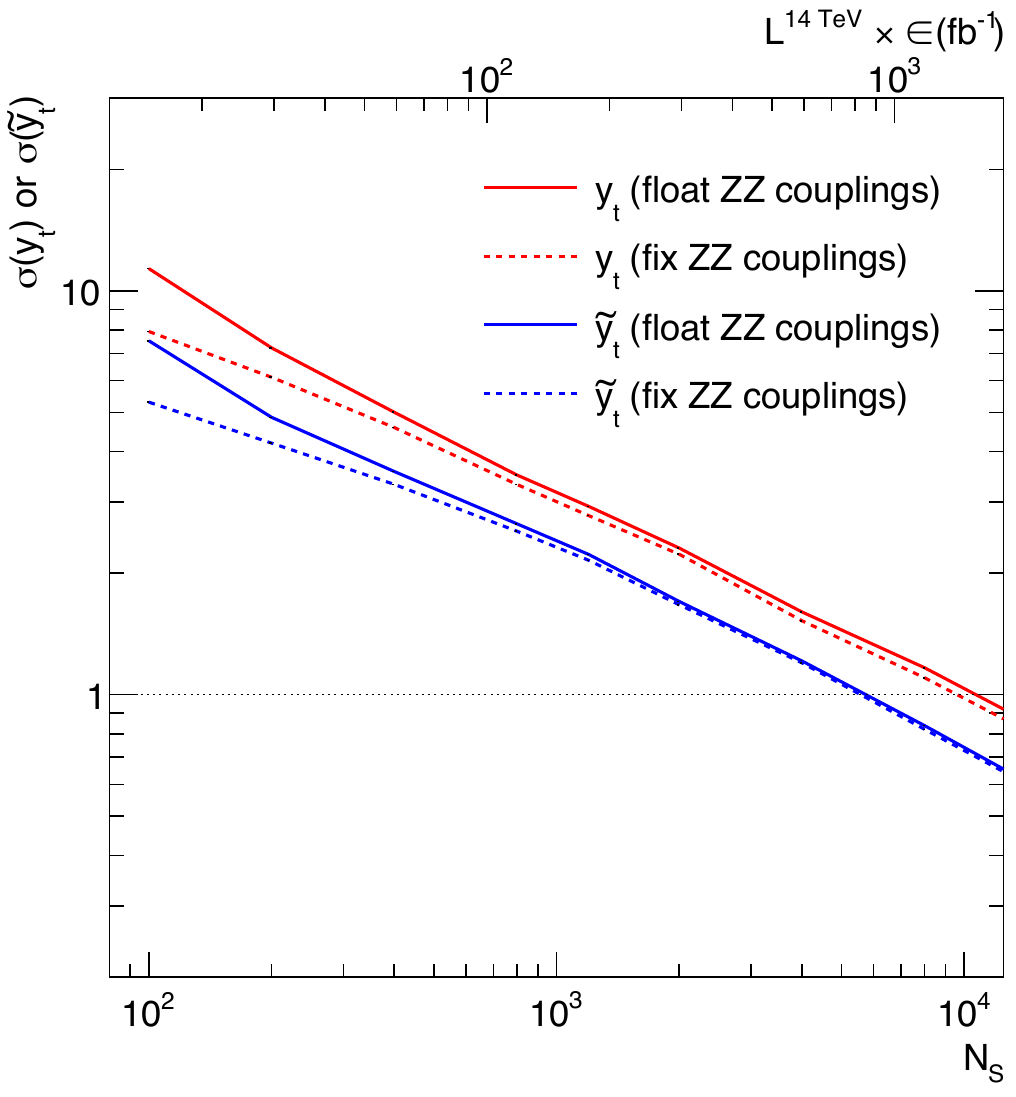}
\end{center}
\caption{Sensitivity curves for $\sigma(y_t)$ (top, red) and $\sigma(\tilde{y}_t)$ (bottom, blue) as function of the number of signal events ($N_S$) (bottom axis) and luminosity $\times$ efficiency (top axis) assuming SM production ($gg\to h$ plus VBF at 14 TeV) and branching ratios~\cite{Dittmaier:2011ti,Heinemeyer:2013tqa}.~In these fits we have utilized the Relaxed$-\Upsilon$ cuts discussed in the text and included both the $h\to4\ell$ ($4\ell \equiv 2e2\mu, 4e, 4\mu$) signal and the $q\bar{q}\to 4\ell$ background.~We fit to a `true' point of $\vec{\lambda} = (1,0 | 0.01,0,0.007)$ corresponding to the SM prediction for the top Yukawa which is indicated by the dotted black line.~We also demonstrate the effect of floating (solid) the effective $ZZ$ couplings (see~\eref{ZZlag}) versus keeping them fixed (dashed).}
\label{fig:curves}
\end{figure}

The crucial point to emphasize is that we should be able to probe $\mathcal{O}(1)$ values of the top Yukawa coupling with $\sim 6000 - 10000$ events corresponding to $\sim 800 - 1500$ fb$^{-1}$ assuming $100\%$ efficiency.~Of course in reality the efficiency is significantly less, so more realistically $\sim 2000 - 5000$ fb$^{-1}$ may be needed depending on detector performance as well production uncertainties.~The lower ends of this range should be within reach at the high-luminosity LHC, and even better sensitivity would be achieved with a future hadron collider at higher energy. 

\subsection{Probing top Yukawa $CP$ Properties}

The results in~\fref{curves} indicate that the LHC or a future collider may be able to directly probe the $CP$ properties of the top Yukawa coupling in $h\to4\ell$.~To further investigate this we show in~\fref{yvya1} and~\fref{yvya2} results from the fit for the $1\sigma$ allowed region in the $y_t - \tilde{y}_t$ plane for a range of data set sizes.~The allowed parameter space corresponds to the entire region inside the ellipse. 
\begin{figure*}[tbh]
\begin{center}
\includegraphics[width=.45\textwidth]{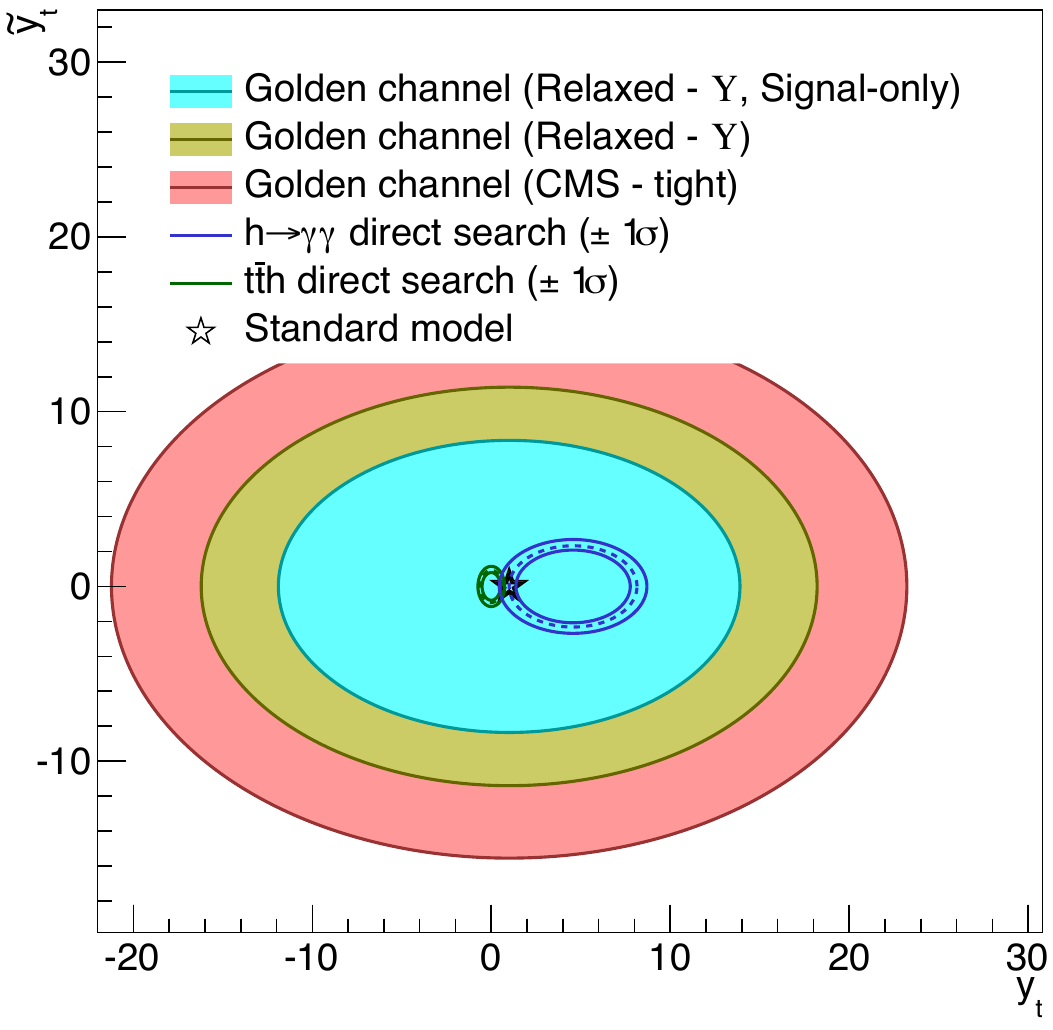}
\includegraphics[width=.45\textwidth]{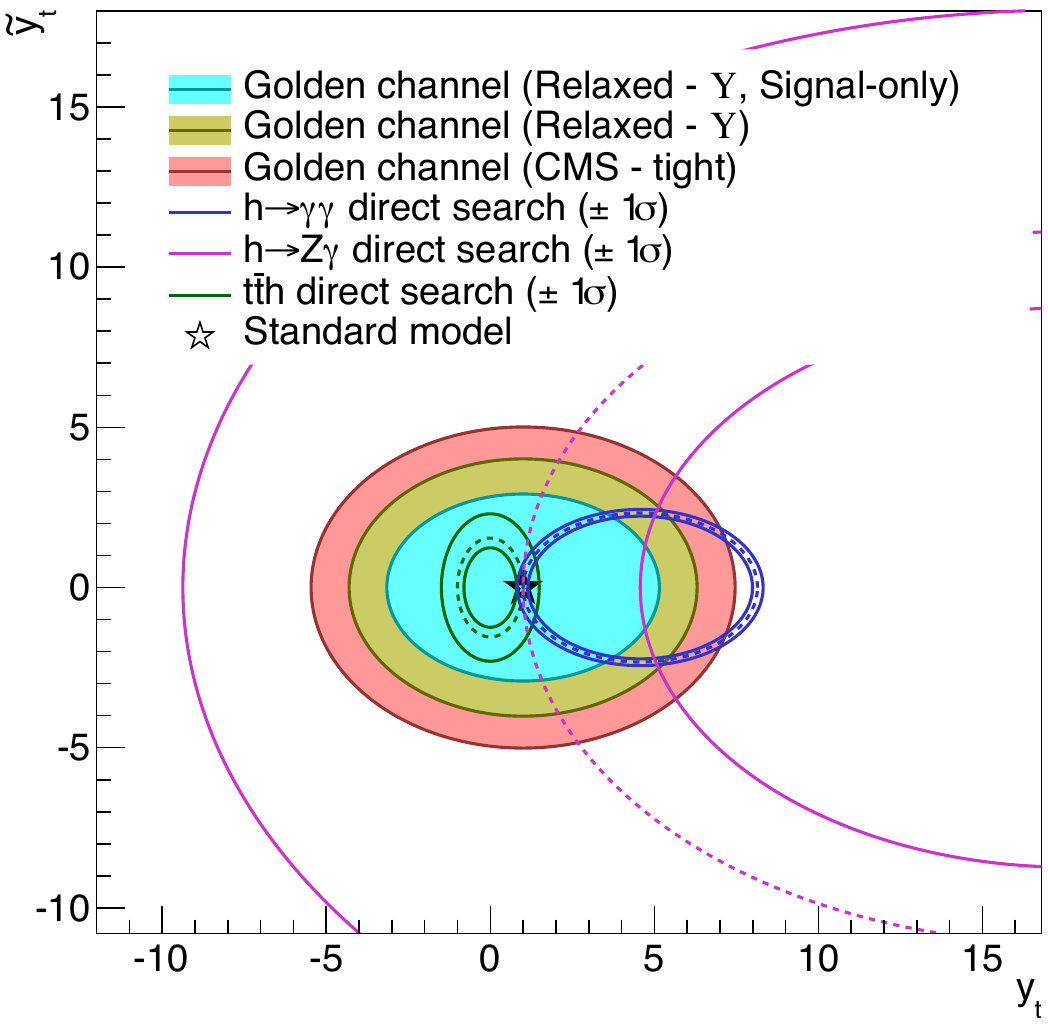}
\end{center}
\caption{{\bf Left:}~$1\sigma$ contours for $y_t$ vs.~$\tilde{y}_t$ with 100 $h\to4\ell$ events corresponding to $\sim 15-40$ fb$^{-1}$ at the LHC14 assuming SM production and branding fractions~\cite{Dittmaier:2011ti,Heinemeyer:2013tqa} and depending on detector efficiencies.~The allowed parameter space is the entire region inside the ellipses.~The same fit as in~\fref{curves} with floating $ZZ$ couplings is performed with the true point represented by the star and corresponding roughly to the SM prediction.~We show the $1\sigma$ confidence interval obtained in $h\to4\ell$ utilizing CMS-like cuts~\cite{CMS-PAS-HIG-14-014,Khachatryan:2014kca} (large, red ellipse) and compare it to the Relaxed$-\Upsilon$ cuts (middle, yellow ellipse) described in text and introduced in~\cite{Chen:2014gka}.~For comparison with the ideal case we also show the projected $1\sigma$ interval assuming a pure signal sample (small, turquoise ellipse) and utilizing the Relaxed$-\Upsilon$ cuts.~The current $1\sigma$ confidence intervals obtained in $tth$ (green band on the left)~\cite{Khachatryan:2014qaa} and $h\to\gamma\gamma$ (blue band on the right)~\cite{Khachatryan:2014ira} direct searches are also shown (see~\tref{other}).~{\bf Right:}~Same as left, but for 800 $h\to4\ell$ events corresponding to $\sim 100-300$ fb$^{-1}$.~The projected $1\sigma$ intervals from $tth$ and $h\to\gamma\gamma$ searches have been used assuming 300 fb$^{-1}$~\cite{Dawson:2013bba,ATLAS-collaboration:1484890}.~We have also added the $1\sigma$ projections from $h\to Z\gamma$ (thick pink band)~\cite{htoAA} searches which start to become relevant at this luminosity.}
\label{fig:yvya1}
\end{figure*}

\begin{figure*}[tbh]
\begin{center}
\includegraphics[width=.45\textwidth]{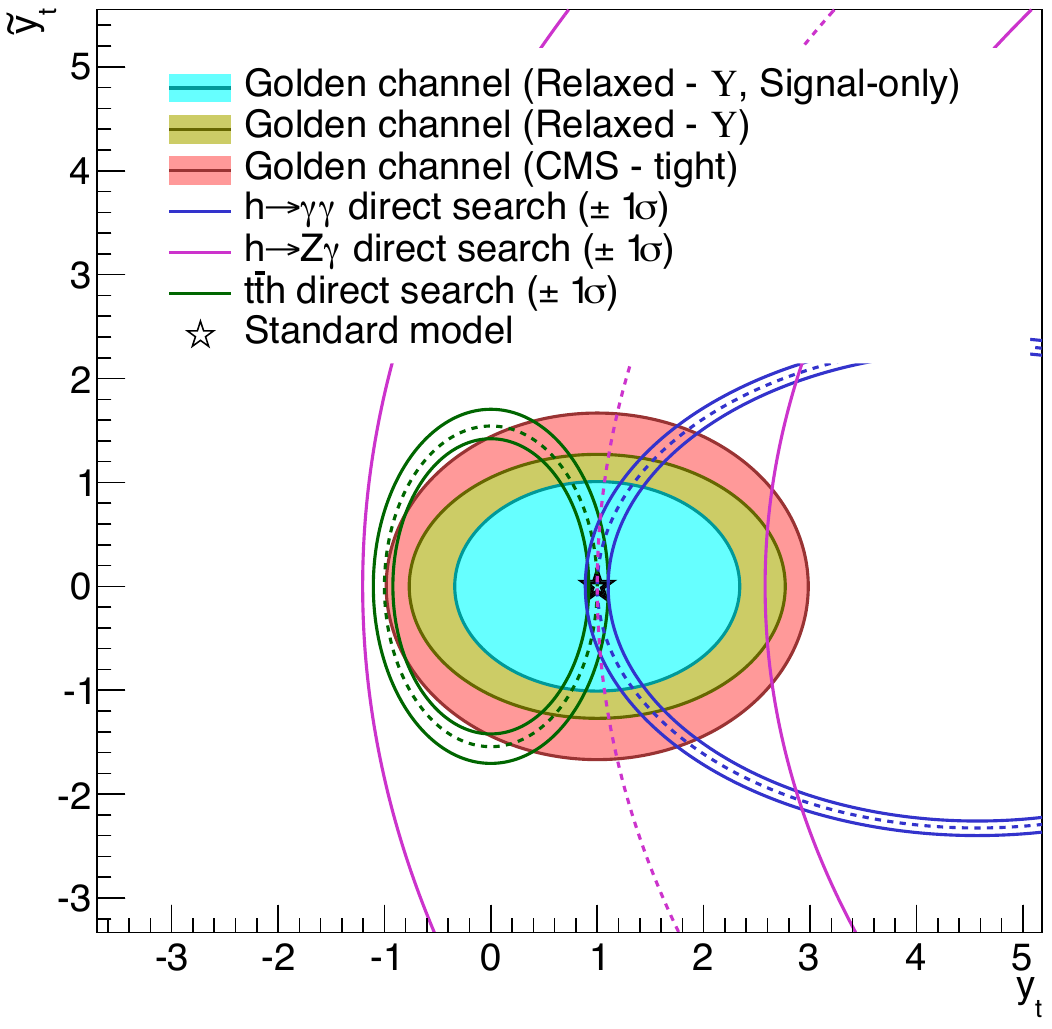}
\includegraphics[width=.45\textwidth]{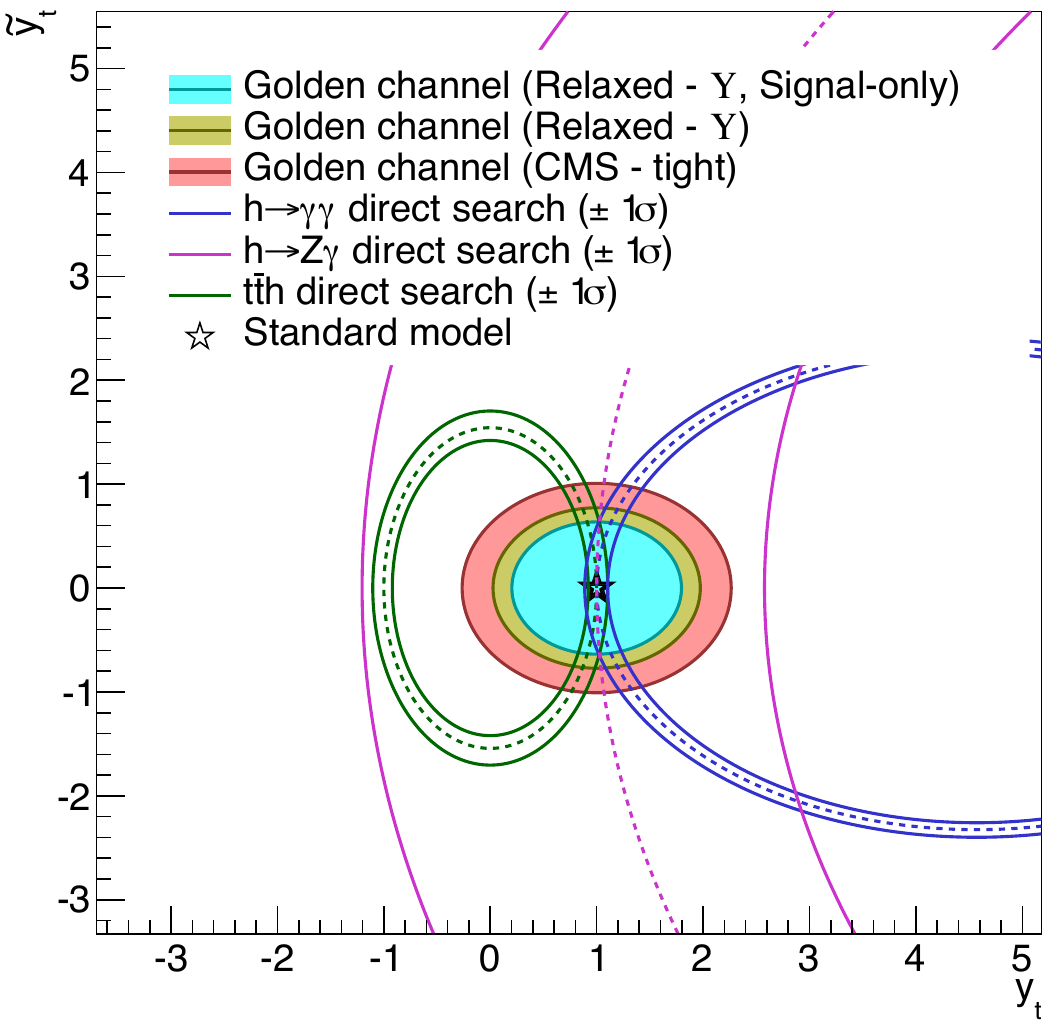}
\end{center}
\caption{{\bf Left:}~Same as~\fref{yvya1}, but for 8000 $h\to4\ell$ events corresponding to $\sim 1000-3000$ fb$^{-1}$ depending on detector efficiencies.~{\bf Right:}~Same as left, but for 20k events corresponding to $\gtrsim 3000$ fb$^{-1}$.~For both plots, the projected $1\sigma$ intervals from $tth$, $h\to \gamma\gamma$, and $h\to Z\gamma$ searches have been used assuming 3000 fb$^{-1}$~\cite{Dawson:2013bba,ATLAS-collaboration:1484890,htoAA} (see~\tref{other}).}
\label{fig:yvya2}
\end{figure*}

In addition to utilizing the Relaxed$-\Upsilon$ cuts (middle, yellow ellipses) as in~\fref{curves}, we also show results using CMS-like cuts~\cite{CMS-PAS-HIG-14-014,Khachatryan:2014kca} (large, red ellipses).~This makes it clear the improved sensitivity obtained when the Relaxed$-\Upsilon$ cuts are used.~For comparison and as a demonstration of the ideal case, we also show the $1\sigma$ region obtained assuming a pure signal sample (inner, turquoise ellipses) using these optimized cuts.~This also makes clear the effects of the $q\bar{q}\to4\ell$ background.

\fref{yvya1} and \fref{yvya2}~also compare the golden channel to other measurements which are sensitive to the top Yukawa coupling:~the $tth$ cross section, the branching ratio of $h\rightarrow \gamma\gamma$, and the branching ratio of $h\rightarrow Z\gamma$.~The $1-\sigma$ contours are derived from the relative signal strength ($\mu_i = \sigma/\sigma_{SM}$ or BR/BR$_{SM}$) for each measurement given by,
\bea
\label{eq:mu}
\mu(tth) &\simeq&
y^2 + 0.42\, \tilde{y}^2 \\
\mu(h\rightarrow \gamma \gamma) &\simeq& 
\left( 1.28-0.28 \, y \right )^2 + \, (0.43 \, \tilde{y})^2  \nonumber\\
\mu(h\rightarrow Z \gamma) &\simeq& 
\left( 1.06-0.06 \, y \right )^2 + \, (0.09 \, \tilde{y})^2 , \nonumber
\eea
where for $\mu(tth)$ we use the cross section at 14 TeV for the approximate value in terms of $y$ and $\tilde{y}$~\cite{ATLAS-collaboration:1484890} and the numerical factors in $h\to V\gamma$ are obtained by evaluating the top and $W$ loops~\cite{Djouadi:2005gi,Djouadi:2005gj} at 125 GeV.~The values we use for the $\mu_i$ signal strengths are summarized in \tref{other}.
\begin{table}[tb]
\centering
\begin{tabular}{|c|c|c|c|}
\hline
$\mathcal{L}$ & $\mu(tth)$ & $\mu(h\rightarrow \gamma\gamma)$  & $\mu(h\rightarrow Z\gamma)$ \\ \hline 
Current & $2.8 \pm 1.0$~\cite{Khachatryan:2014qaa} & $1.14 \pm 0.25$~\cite{Khachatryan:2014ira}  &  NA \\ \hline
300 fb$^{-1}$ & $1.0 \pm 0.55$~\cite{ATLAS-collaboration:1484890} & $1.0 \pm 0.1$~\cite{Dawson:2013bba}& $1.0 \pm 0.6$~\cite{htoAA} \\ \hline
3000 fb$^{-1}$ & $1.0 \pm 0.18$~\cite{ATLAS-collaboration:1484890} & $1.0 \pm 0.05$~\cite{Dawson:2013bba}  & $1.0 \pm 0.2$~\cite{htoAA} \\ \hline
\end{tabular}
\caption{Values of current constraints and future projections on the relative signal strength $\mu_i = \sigma/\sigma_{SM}$ (or $BR/BR_{SM}$) for given luminosities.}
\label{tab:other}
\end{table}

Before discussing our results further, we comment that from the numerical values in~\eref{mu}, it is clear that the sensitivity to the top Yukawa in $h\to4\ell$ is driven by the $\gamma\gamma$ intermediate states.~This implies that a reasonable approximation of the sensitivity to $y_t$ and $\tilde{y}_t$ could have simply been obtained from a naive rescaling of the results for the sensitivity to the $\gamma\gamma$ effective operators found in~\cite{Chen:2014gka,Chen:2015iha}.~However, we emphasize that this rescaling ignores potential correlations between the $Z\gamma$ and $\gamma\gamma$ effective operators~\cite{Chen:2012jy,Chen:2013ejz,Chen:2014pia}.~Furthermore, the parameter fitting done in this study is qualitatively different since (ignoring $ZZ$ couplings) only two parameters ($y_t, \tilde{y}_t$) are floated in contrast to four ($A_2^{Z\gamma}, A_3^{Z\gamma}, A_2^{\gamma\gamma}, A_2^{\gamma\gamma}$) when using effective couplings.~For these reasons we have not simply done a rescaling of the effective couplings, though the end results for the sensitivity to $y_t$ and $\tilde{y}_t$ are not drastically different.

The current $1\sigma$ confidence intervals obtained in $tth$ (green band on the left)~\cite{Khachatryan:2014qaa} and $h\to\gamma\gamma$ (blue band on the right)~\cite{Khachatryan:2014ira} direct searches are shown on the left in~\fref{yvya1} where 100 $h\to 4\ell$ events have been assumed.~We see that at this stage $h\to4\ell$ is not competitive with $tth$ and $h\to\gamma\gamma$ searches.~For 800 events shown on the right we use the projected $1\sigma$ intervals from $tth$ and $h\to\gamma\gamma$ searches assuming 300 fb$^{-1}$~\cite{Dawson:2013bba,ATLAS-collaboration:1484890} and a SM-like central value.~We have also added the $1\sigma$ projections from $h\to Z\gamma$ (thick pink band)~\cite{htoAA} searches which start to become relevant at this luminosity.~We can see at this stage that $h\to4\ell$ is also starting to become a useful channel to complement $tth$ and $h\to V\gamma$ searches for studying the top Yukawa.

In~\fref{yvya2} we show the same results, but for 8000 (left) and 20k (right) events corresponding to $\gtrsim 1000-3000$ fb$^{-1}$ and where the projected $1\sigma$ intervals from $tth$, $h\to \gamma\gamma$, and $h\to Z\gamma$ searches have been used assuming 3000 fb$^{-1}$~\cite{Dawson:2013bba,ATLAS-collaboration:1484890,htoAA}.~We see in these results that if we assume the Higgs couplings to $ZZ$ and $WW$ are positive, eventually $h\to4\ell$ should be able to establish the overall sign of $y_t$ independently of any other measurements of the top Yukawa.~We further see the possibility of using $h\to4\ell$ as a consistency check with $tth$ and $h\to V\gamma$ searches as well as the qualitatively different nature of the $h\to4\ell$ measurement.

The results in~\fref{yvya1} and \fref{yvya2} make it clear that $h\to4\ell$ is a useful and complementary channel to $tth$, $h\to Z\gamma$, and $h\to\gamma\gamma$ searches for probing the top Yukawa at the LHC or a future collider.~Furthermore, depending on how sensitivities evolve over time, it may be possible that $h\to4\ell$ will be able to constrain regions of parameter space which are difficult to probe in other channels helping to ensure that potential $CP$ violating effects would not go unnoticed.~In the event where a deviation from the SM value is observed in either on-shell $h\rightarrow Z\gamma, \gamma\gamma$ two body decays or $tth$ production, the four lepton channel will be a crucial ingredient in both confirming and characterizing the anomaly.~Quantifying more precisely these possibilities will require a detailed treatment of the various one-loop and off-shell effects which we have not included, but a thorough investigation is left to ongoing work~\cite{followup2}.~Many more results from the current analysis can be found in~\cite{WEBSITE}.

\section{Conclusions}
\label{sec:conclution}

We have demonstrated that the $h\to 4\ell$ `golden channel' can be a useful probe of the top Yukawa at the LHC and future colliders.~We have considered the leading effects in order to give a proof of principle that this channel can serve as a complementary, but qualitatively different, measurement to $h\to\gamma\gamma$ and $h\to Z\gamma$ two body decays as well as $gg\to h$ and $tth$ searches for studying the top Yukawa.~A detailed study of the sub-dominant one-loop and off-shell effects in order to quantify the sensitivity to the top Yukawa more precisely is ongoing.

In particular the $h\to4\ell$ channel can be used to directly study the $CP$ properties of the top Yukawa in a single channel independent of other measurements.~This is useful because multiple measurements need not be combined allowing us to avoid complications from combining errors in different channels in order to establish the $CP$ properties.~Furthermore, the experimentally clean nature and high precision with which this channel is measured along with the fact that it is theoretically very well understood makes it valuable as both a consistency check for other channels as well as perhaps the most direct way to uncover potential $CP$ violation in the top Yukawa.

The main drawback of $h\to4\ell$ is that it is statistics limited, but our results indicate that the necessary precision to begin probing the top Yukawa may be reached at the LHC and certainly at a future hadron collider.~The theoretical importance of the top Yukawa coupling has been firmly established for quite some time and finding as many independent probes to study it will be crucial.~We thus encourage experimentalists to add $h\to4\ell$ to the list of already established channels for studying the top Yukawa and in particular its $CP$ properties.

~\\
\noindent
{\bf Acknowledgments:}~We thank Maria Spiropulu for providing us with the resources necessary to complete this study as well as Simon Badger, Fabrizio Caola, Adam Falkowski, Gian Giudice, Roni Harnik, Joe Lykken, Tom Melia, Markus Schulze, Pedro Schwaller, and Tim Tait for comments and helpful discussions.~R.V.M.~is supported by the ERC Advanced Grant Higgs@LHC.~Y.C.~is supported by the Weston Havens Foundation and DOE grant No.~DE-FG02-92-ER-40701.~D.S.~and~R.V.M.~would also like to thank the participants of the workshop ``After the Discovery: Hunting for a Non-Standard Higgs Sector" at Centro de Ciencias de Benasque Pedro Pascual for lively atmosphere and discussions where this work began.


\bibliographystyle{apsrev}
\bibliography{references}

\begin{thebibliography}{104}
\expandafter\ifx\csname natexlab\endcsname\relax\def\natexlab#1{#1}\fi
\expandafter\ifx\csname bibnamefont\endcsname\relax
  \def\bibnamefont#1{#1}\fi
\expandafter\ifx\csname bibfnamefont\endcsname\relax
  \def\bibfnamefont#1{#1}\fi
\expandafter\ifx\csname citenamefont\endcsname\relax
  \def\citenamefont#1{#1}\fi
\expandafter\ifx\csname url\endcsname\relax
  \def\url#1{\texttt{#1}}\fi
\expandafter\ifx\csname urlprefix\endcsname\relax\def\urlprefix{URL }\fi
\providecommand{\bibinfo}[2]{#2}
\providecommand{\eprint}[2][]{\url{#2}}

\bibitem[{\citenamefont{Aad et~al.}(2012)}]{:2012gk}
\bibinfo{author}{\bibfnamefont{G.}~\bibnamefont{Aad}} \bibnamefont{et~al.}
  (\bibinfo{collaboration}{ATLAS Collaboration}), \bibinfo{journal}{Phys.Lett.}
  \textbf{\bibinfo{volume}{B716}}, \bibinfo{pages}{1} (\bibinfo{year}{2012}),
  \eprint{1207.7214}.

\bibitem[{\citenamefont{Chatrchyan et~al.}(2012)}]{:2012gu}
\bibinfo{author}{\bibfnamefont{S.}~\bibnamefont{Chatrchyan}}
  \bibnamefont{et~al.} (\bibinfo{collaboration}{CMS Collaboration}),
  \bibinfo{journal}{Phys.Lett.} \textbf{\bibinfo{volume}{B716}},
  \bibinfo{pages}{30} (\bibinfo{year}{2012}), \eprint{1207.7235}.

\bibitem[{\citenamefont{Aad et~al.}(2013)}]{Aad:2013xqa}
\bibinfo{author}{\bibfnamefont{G.}~\bibnamefont{Aad}} \bibnamefont{et~al.}
  (\bibinfo{collaboration}{ATLAS Collaboration}), \bibinfo{journal}{Phys.Lett.}
  \textbf{\bibinfo{volume}{B726}}, \bibinfo{pages}{120} (\bibinfo{year}{2013}),
  \eprint{1307.1432}.

\bibitem[{\citenamefont{Chatrchyan
  et~al.}(2014{\natexlab{a}})}]{Chatrchyan:2013mxa}
\bibinfo{author}{\bibfnamefont{S.}~\bibnamefont{Chatrchyan}}
  \bibnamefont{et~al.} (\bibinfo{collaboration}{CMS Collaboration}),
  \bibinfo{journal}{Phys.Rev.} \textbf{\bibinfo{volume}{D89}},
  \bibinfo{pages}{092007} (\bibinfo{year}{2014}{\natexlab{a}}),
  \eprint{1312.5353}.

\bibitem[{\citenamefont{Khachatryan
  et~al.}(2014{\natexlab{a}})}]{Khachatryan:2014qaa}
\bibinfo{author}{\bibfnamefont{V.}~\bibnamefont{Khachatryan}}
  \bibnamefont{et~al.} (\bibinfo{collaboration}{CMS}), \bibinfo{journal}{JHEP}
  \textbf{\bibinfo{volume}{1409}}, \bibinfo{pages}{087}
  (\bibinfo{year}{2014}{\natexlab{a}}), \eprint{1408.1682}.

\bibitem[{ATL(2014{\natexlab{a}})}]{ATLAS-CONF-2014-011}
\bibinfo{type}{Tech. Rep.} \bibinfo{number}{ATLAS-CONF-2014-011},
  \bibinfo{institution}{CERN}, \bibinfo{address}{Geneva}
  (\bibinfo{year}{2014}{\natexlab{a}}).

\bibitem[{\citenamefont{Aad et~al.}(2014)}]{Aad:2014lma}
\bibinfo{author}{\bibfnamefont{G.}~\bibnamefont{Aad}} \bibnamefont{et~al.}
  (\bibinfo{collaboration}{ATLAS Collaboration}), \bibinfo{journal}{Phys.Lett.}
  \textbf{\bibinfo{volume}{B740}}, \bibinfo{pages}{222} (\bibinfo{year}{2014}),
  \eprint{1409.3122}.

\bibitem[{\citenamefont{Gunion and He}(1996)}]{Gunion:1996xu}
\bibinfo{author}{\bibfnamefont{J.~F.} \bibnamefont{Gunion}} \bibnamefont{and}
  \bibinfo{author}{\bibfnamefont{X.-G.} \bibnamefont{He}},
  \bibinfo{journal}{Phys.Rev.Lett.} \textbf{\bibinfo{volume}{76}},
  \bibinfo{pages}{4468} (\bibinfo{year}{1996}), \eprint{hep-ph/9602226}.

\bibitem[{\citenamefont{Gunion and Pliszka}(1998)}]{Gunion:1998hm}
\bibinfo{author}{\bibfnamefont{J.~F.} \bibnamefont{Gunion}} \bibnamefont{and}
  \bibinfo{author}{\bibfnamefont{J.}~\bibnamefont{Pliszka}},
  \bibinfo{journal}{Phys.Lett.} \textbf{\bibinfo{volume}{B444}},
  \bibinfo{pages}{136} (\bibinfo{year}{1998}), \eprint{hep-ph/9809306}.

\bibitem[{\citenamefont{Ellis et~al.}(2014)\citenamefont{Ellis, Hwang, Sakurai,
  and Takeuchi}}]{Ellis:2013yxa}
\bibinfo{author}{\bibfnamefont{J.}~\bibnamefont{Ellis}},
  \bibinfo{author}{\bibfnamefont{D.~S.} \bibnamefont{Hwang}},
  \bibinfo{author}{\bibfnamefont{K.}~\bibnamefont{Sakurai}}, \bibnamefont{and}
  \bibinfo{author}{\bibfnamefont{M.}~\bibnamefont{Takeuchi}},
  \bibinfo{journal}{JHEP} \textbf{\bibinfo{volume}{1404}}, \bibinfo{pages}{004}
  (\bibinfo{year}{2014}), \eprint{1312.5736}.

\bibitem[{\citenamefont{Khatibi and Najafabadi}(2014)}]{Khatibi:2014bsa}
\bibinfo{author}{\bibfnamefont{S.}~\bibnamefont{Khatibi}} \bibnamefont{and}
  \bibinfo{author}{\bibfnamefont{M.~M.} \bibnamefont{Najafabadi}},
  \bibinfo{journal}{Phys.Rev.} \textbf{\bibinfo{volume}{D90}},
  \bibinfo{pages}{074014} (\bibinfo{year}{2014}), \eprint{1409.6553}.

\bibitem[{\citenamefont{He et~al.}(2014)\citenamefont{He, Li, and
  Zheng}}]{He:2014xla}
\bibinfo{author}{\bibfnamefont{X.-G.} \bibnamefont{He}},
  \bibinfo{author}{\bibfnamefont{G.-N.} \bibnamefont{Li}}, \bibnamefont{and}
  \bibinfo{author}{\bibfnamefont{Y.-J.} \bibnamefont{Zheng}}
  (\bibinfo{year}{2014}), \eprint{1501.00012}.

\bibitem[{\citenamefont{Boudjema et~al.}(2015)\citenamefont{Boudjema, Godbole,
  Guadagnoli, and Mohan}}]{Boudjema:2015nda}
\bibinfo{author}{\bibfnamefont{F.}~\bibnamefont{Boudjema}},
  \bibinfo{author}{\bibfnamefont{R.~M.} \bibnamefont{Godbole}},
  \bibinfo{author}{\bibfnamefont{D.}~\bibnamefont{Guadagnoli}},
  \bibnamefont{and} \bibinfo{author}{\bibfnamefont{K.~A.} \bibnamefont{Mohan}}
  (\bibinfo{year}{2015}), \eprint{1501.03157}.

\bibitem[{\citenamefont{Khachatryan
  et~al.}(2014{\natexlab{b}})}]{Khachatryan:2014jba}
\bibinfo{author}{\bibfnamefont{V.}~\bibnamefont{Khachatryan}}
  \bibnamefont{et~al.} (\bibinfo{collaboration}{CMS Collaboration})
  (\bibinfo{year}{2014}{\natexlab{b}}), \eprint{1412.8662}.

\bibitem[{ATL(2014{\natexlab{b}})}]{ATLAS-CONF-2014-010}
\bibinfo{type}{Tech. Rep.} \bibinfo{number}{ATLAS-CONF-2014-010},
  \bibinfo{institution}{CERN}, \bibinfo{address}{Geneva}
  (\bibinfo{year}{2014}{\natexlab{b}}).

\bibitem[{\citenamefont{Carmi et~al.}(2012)\citenamefont{Carmi, Falkowski,
  Kuflik, Volansky, and Zupan}}]{Carmi:2012in}
\bibinfo{author}{\bibfnamefont{D.}~\bibnamefont{Carmi}},
  \bibinfo{author}{\bibfnamefont{A.}~\bibnamefont{Falkowski}},
  \bibinfo{author}{\bibfnamefont{E.}~\bibnamefont{Kuflik}},
  \bibinfo{author}{\bibfnamefont{T.}~\bibnamefont{Volansky}}, \bibnamefont{and}
  \bibinfo{author}{\bibfnamefont{J.}~\bibnamefont{Zupan}},
  \bibinfo{journal}{JHEP} \textbf{\bibinfo{volume}{1210}}, \bibinfo{pages}{196}
  (\bibinfo{year}{2012}), \eprint{1207.1718}.

\bibitem[{\citenamefont{Banerjee et~al.}(2012)\citenamefont{Banerjee,
  Mukhopadhyay, and Mukhopadhyaya}}]{Banerjee:2012xc}
\bibinfo{author}{\bibfnamefont{S.}~\bibnamefont{Banerjee}},
  \bibinfo{author}{\bibfnamefont{S.}~\bibnamefont{Mukhopadhyay}},
  \bibnamefont{and}
  \bibinfo{author}{\bibfnamefont{B.}~\bibnamefont{Mukhopadhyaya}},
  \bibinfo{journal}{JHEP} \textbf{\bibinfo{volume}{1210}}, \bibinfo{pages}{062}
  (\bibinfo{year}{2012}), \eprint{1207.3588}.

\bibitem[{\citenamefont{Plehn and Rauch}(2012)}]{Plehn:2012iz}
\bibinfo{author}{\bibfnamefont{T.}~\bibnamefont{Plehn}} \bibnamefont{and}
  \bibinfo{author}{\bibfnamefont{M.}~\bibnamefont{Rauch}},
  \bibinfo{journal}{Europhys.Lett.} \textbf{\bibinfo{volume}{100}},
  \bibinfo{pages}{11002} (\bibinfo{year}{2012}), \eprint{1207.6108}.

\bibitem[{\citenamefont{Djouadi}(2013)}]{Djouadi:2012rh}
\bibinfo{author}{\bibfnamefont{A.}~\bibnamefont{Djouadi}},
  \bibinfo{journal}{Eur.Phys.J.} \textbf{\bibinfo{volume}{C73}},
  \bibinfo{pages}{2498} (\bibinfo{year}{2013}), \eprint{1208.3436}.

\bibitem[{\citenamefont{Belanger et~al.}(2013)\citenamefont{Belanger, Dumont,
  Ellwanger, Gunion, and Kraml}}]{Belanger:2012gc}
\bibinfo{author}{\bibfnamefont{G.}~\bibnamefont{Belanger}},
  \bibinfo{author}{\bibfnamefont{B.}~\bibnamefont{Dumont}},
  \bibinfo{author}{\bibfnamefont{U.}~\bibnamefont{Ellwanger}},
  \bibinfo{author}{\bibfnamefont{J.}~\bibnamefont{Gunion}}, \bibnamefont{and}
  \bibinfo{author}{\bibfnamefont{S.}~\bibnamefont{Kraml}},
  \bibinfo{journal}{JHEP} \textbf{\bibinfo{volume}{1302}}, \bibinfo{pages}{053}
  (\bibinfo{year}{2013}), \eprint{1212.5244}.

\bibitem[{\citenamefont{Cheung et~al.}(2013)\citenamefont{Cheung, Lee, and
  Tseng}}]{Cheung:2013kla}
\bibinfo{author}{\bibfnamefont{K.}~\bibnamefont{Cheung}},
  \bibinfo{author}{\bibfnamefont{J.~S.} \bibnamefont{Lee}}, \bibnamefont{and}
  \bibinfo{author}{\bibfnamefont{P.-Y.} \bibnamefont{Tseng}},
  \bibinfo{journal}{JHEP} \textbf{\bibinfo{volume}{1305}}, \bibinfo{pages}{134}
  (\bibinfo{year}{2013}), \eprint{1302.3794}.

\bibitem[{\citenamefont{Falkowski et~al.}(2013)\citenamefont{Falkowski, Riva,
  and Urbano}}]{Falkowski:2013dza}
\bibinfo{author}{\bibfnamefont{A.}~\bibnamefont{Falkowski}},
  \bibinfo{author}{\bibfnamefont{F.}~\bibnamefont{Riva}}, \bibnamefont{and}
  \bibinfo{author}{\bibfnamefont{A.}~\bibnamefont{Urbano}}
  (\bibinfo{year}{2013}), \eprint{1303.1812}.

\bibitem[{\citenamefont{Giardino et~al.}(2014)\citenamefont{Giardino, Kannike,
  Masina, Raidal, and Strumia}}]{Giardino:2013bma}
\bibinfo{author}{\bibfnamefont{P.~P.} \bibnamefont{Giardino}},
  \bibinfo{author}{\bibfnamefont{K.}~\bibnamefont{Kannike}},
  \bibinfo{author}{\bibfnamefont{I.}~\bibnamefont{Masina}},
  \bibinfo{author}{\bibfnamefont{M.}~\bibnamefont{Raidal}}, \bibnamefont{and}
  \bibinfo{author}{\bibfnamefont{A.}~\bibnamefont{Strumia}},
  \bibinfo{journal}{JHEP} \textbf{\bibinfo{volume}{1405}}, \bibinfo{pages}{046}
  (\bibinfo{year}{2014}), \eprint{1303.3570}.

\bibitem[{\citenamefont{Ellis and You}(2013)}]{Ellis:2013lra}
\bibinfo{author}{\bibfnamefont{J.}~\bibnamefont{Ellis}} \bibnamefont{and}
  \bibinfo{author}{\bibfnamefont{T.}~\bibnamefont{You}},
  \bibinfo{journal}{JHEP} \textbf{\bibinfo{volume}{1306}}, \bibinfo{pages}{103}
  (\bibinfo{year}{2013}), \eprint{1303.3879}.

\bibitem[{\citenamefont{Bernon et~al.}(2014)\citenamefont{Bernon, Dumont, and
  Kraml}}]{Bernon:2014vta}
\bibinfo{author}{\bibfnamefont{J.}~\bibnamefont{Bernon}},
  \bibinfo{author}{\bibfnamefont{B.}~\bibnamefont{Dumont}}, \bibnamefont{and}
  \bibinfo{author}{\bibfnamefont{S.}~\bibnamefont{Kraml}},
  \bibinfo{journal}{Phys.Rev.} \textbf{\bibinfo{volume}{D90}},
  \bibinfo{pages}{071301} (\bibinfo{year}{2014}), \eprint{1409.1588}.

\bibitem[{\citenamefont{Tait and Yuan}(2000)}]{Tait:2000sh}
\bibinfo{author}{\bibfnamefont{T.~M.} \bibnamefont{Tait}} \bibnamefont{and}
  \bibinfo{author}{\bibfnamefont{C.-P.} \bibnamefont{Yuan}},
  \bibinfo{journal}{Phys.Rev.} \textbf{\bibinfo{volume}{D63}},
  \bibinfo{pages}{014018} (\bibinfo{year}{2000}), \eprint{hep-ph/0007298}.

\bibitem[{\citenamefont{Agashe et~al.}(2013)}]{Agashe:2013hma}
\bibinfo{author}{\bibfnamefont{K.}~\bibnamefont{Agashe}} \bibnamefont{et~al.}
  (\bibinfo{collaboration}{Top Quark Working Group}) (\bibinfo{year}{2013}),
  \eprint{1311.2028}.

\bibitem[{\citenamefont{Chang et~al.}(2014)\citenamefont{Chang, Cheung, Lee,
  and Lu}}]{Chang:2014rfa}
\bibinfo{author}{\bibfnamefont{J.}~\bibnamefont{Chang}},
  \bibinfo{author}{\bibfnamefont{K.}~\bibnamefont{Cheung}},
  \bibinfo{author}{\bibfnamefont{J.~S.} \bibnamefont{Lee}}, \bibnamefont{and}
  \bibinfo{author}{\bibfnamefont{C.-T.} \bibnamefont{Lu}},
  \bibinfo{journal}{JHEP} \textbf{\bibinfo{volume}{1405}}, \bibinfo{pages}{062}
  (\bibinfo{year}{2014}), \eprint{1403.2053}.

\bibitem[{\citenamefont{Yue}(2014)}]{Yue:2014tya}
\bibinfo{author}{\bibfnamefont{J.}~\bibnamefont{Yue}} (\bibinfo{year}{2014}),
  \eprint{1410.2701}.

\bibitem[{\citenamefont{Demartin et~al.}(2015)\citenamefont{Demartin, Maltoni,
  Mawatari, and Zaro}}]{Demartin:2015uha}
\bibinfo{author}{\bibfnamefont{F.}~\bibnamefont{Demartin}},
  \bibinfo{author}{\bibfnamefont{F.}~\bibnamefont{Maltoni}},
  \bibinfo{author}{\bibfnamefont{K.}~\bibnamefont{Mawatari}}, \bibnamefont{and}
  \bibinfo{author}{\bibfnamefont{M.}~\bibnamefont{Zaro}}
  (\bibinfo{year}{2015}), \eprint{1504.00611}.

\bibitem[{\citenamefont{Grojean et~al.}(2014)\citenamefont{Grojean, Salvioni,
  Schlaffer, and Weiler}}]{Grojean:2013nya}
\bibinfo{author}{\bibfnamefont{C.}~\bibnamefont{Grojean}},
  \bibinfo{author}{\bibfnamefont{E.}~\bibnamefont{Salvioni}},
  \bibinfo{author}{\bibfnamefont{M.}~\bibnamefont{Schlaffer}},
  \bibnamefont{and} \bibinfo{author}{\bibfnamefont{A.}~\bibnamefont{Weiler}},
  \bibinfo{journal}{JHEP} \textbf{\bibinfo{volume}{1405}}, \bibinfo{pages}{022}
  (\bibinfo{year}{2014}), \eprint{1312.3317}.

\bibitem[{\citenamefont{Demartin et~al.}(2014)\citenamefont{Demartin, Maltoni,
  Mawatari, Page, and Zaro}}]{Demartin:2014fia}
\bibinfo{author}{\bibfnamefont{F.}~\bibnamefont{Demartin}},
  \bibinfo{author}{\bibfnamefont{F.}~\bibnamefont{Maltoni}},
  \bibinfo{author}{\bibfnamefont{K.}~\bibnamefont{Mawatari}},
  \bibinfo{author}{\bibfnamefont{B.}~\bibnamefont{Page}}, \bibnamefont{and}
  \bibinfo{author}{\bibfnamefont{M.}~\bibnamefont{Zaro}},
  \bibinfo{journal}{Eur.Phys.J.} \textbf{\bibinfo{volume}{C74}},
  \bibinfo{pages}{3065} (\bibinfo{year}{2014}), \eprint{1407.5089}.

\bibitem[{\citenamefont{Brod et~al.}(2013)\citenamefont{Brod, Haisch, and
  Zupan}}]{Brod:2013cka}
\bibinfo{author}{\bibfnamefont{J.}~\bibnamefont{Brod}},
  \bibinfo{author}{\bibfnamefont{U.}~\bibnamefont{Haisch}}, \bibnamefont{and}
  \bibinfo{author}{\bibfnamefont{J.}~\bibnamefont{Zupan}},
  \bibinfo{journal}{JHEP} \textbf{\bibinfo{volume}{1311}}, \bibinfo{pages}{180}
  (\bibinfo{year}{2013}), \eprint{1310.1385}.

\bibitem[{\citenamefont{Nelson}(1988)}]{Nelson:1986ki}
\bibinfo{author}{\bibfnamefont{C.~A.} \bibnamefont{Nelson}},
  \bibinfo{journal}{Phys.Rev.} \textbf{\bibinfo{volume}{D37}},
  \bibinfo{pages}{1220} (\bibinfo{year}{1988}).

\bibitem[{\citenamefont{Soni and Xu}(1993)}]{Soni:1993jc}
\bibinfo{author}{\bibfnamefont{A.}~\bibnamefont{Soni}} \bibnamefont{and}
  \bibinfo{author}{\bibfnamefont{R.}~\bibnamefont{Xu}},
  \bibinfo{journal}{Phys.Rev.} \textbf{\bibinfo{volume}{D48}},
  \bibinfo{pages}{5259} (\bibinfo{year}{1993}), \eprint{hep-ph/9301225}.

\bibitem[{\citenamefont{Chang et~al.}(1993)\citenamefont{Chang, Keung, and
  Phillips}}]{Chang:1993jy}
\bibinfo{author}{\bibfnamefont{D.}~\bibnamefont{Chang}},
  \bibinfo{author}{\bibfnamefont{W.-Y.} \bibnamefont{Keung}}, \bibnamefont{and}
  \bibinfo{author}{\bibfnamefont{I.}~\bibnamefont{Phillips}},
  \bibinfo{journal}{Phys.Rev.} \textbf{\bibinfo{volume}{D48}},
  \bibinfo{pages}{3225} (\bibinfo{year}{1993}), \eprint{hep-ph/9303226}.

\bibitem[{\citenamefont{Barger et~al.}(1994)\citenamefont{Barger, Cheung,
  Djouadi, Kniehl, and Zerwas}}]{Barger:1993wt}
\bibinfo{author}{\bibfnamefont{V.~D.} \bibnamefont{Barger}},
  \bibinfo{author}{\bibfnamefont{K.-m.} \bibnamefont{Cheung}},
  \bibinfo{author}{\bibfnamefont{A.}~\bibnamefont{Djouadi}},
  \bibinfo{author}{\bibfnamefont{B.~A.} \bibnamefont{Kniehl}},
  \bibnamefont{and} \bibinfo{author}{\bibfnamefont{P.}~\bibnamefont{Zerwas}},
  \bibinfo{journal}{Phys.Rev.} \textbf{\bibinfo{volume}{D49}},
  \bibinfo{pages}{79} (\bibinfo{year}{1994}), \eprint{hep-ph/9306270}.

\bibitem[{\citenamefont{Arens and Sehgal}(1995)}]{Arens:1994wd}
\bibinfo{author}{\bibfnamefont{T.}~\bibnamefont{Arens}} \bibnamefont{and}
  \bibinfo{author}{\bibfnamefont{L.}~\bibnamefont{Sehgal}},
  \bibinfo{journal}{Z.Phys.} \textbf{\bibinfo{volume}{C66}},
  \bibinfo{pages}{89} (\bibinfo{year}{1995}), \eprint{hep-ph/9409396}.

\bibitem[{\citenamefont{Choi et~al.}(2003)\citenamefont{Choi, Miller,
  Muhlleitner, and Zerwas}}]{Choi:2002jk}
\bibinfo{author}{\bibfnamefont{S.}~\bibnamefont{Choi}},
  \bibinfo{author}{\bibfnamefont{.}~\bibnamefont{Miller}, \bibfnamefont{D.J.}},
  \bibinfo{author}{\bibfnamefont{M.}~\bibnamefont{Muhlleitner}},
  \bibnamefont{and} \bibinfo{author}{\bibfnamefont{P.}~\bibnamefont{Zerwas}},
  \bibinfo{journal}{Phys.Lett.} \textbf{\bibinfo{volume}{B553}},
  \bibinfo{pages}{61} (\bibinfo{year}{2003}), \eprint{hep-ph/0210077}.

\bibitem[{\citenamefont{Buszello et~al.}(2004)\citenamefont{Buszello, Fleck,
  Marquard, and van~der Bij}}]{Buszello:2002uu}
\bibinfo{author}{\bibfnamefont{C.}~\bibnamefont{Buszello}},
  \bibinfo{author}{\bibfnamefont{I.}~\bibnamefont{Fleck}},
  \bibinfo{author}{\bibfnamefont{P.}~\bibnamefont{Marquard}}, \bibnamefont{and}
  \bibinfo{author}{\bibfnamefont{J.}~\bibnamefont{van~der Bij}},
  \bibinfo{journal}{Eur.Phys.J.} \textbf{\bibinfo{volume}{C32}},
  \bibinfo{pages}{209} (\bibinfo{year}{2004}), \eprint{hep-ph/0212396}.

\bibitem[{\citenamefont{Godbole et~al.}(2007)\citenamefont{Godbole, Miller, and
  Muhlleitner}}]{Godbole:2007cn}
\bibinfo{author}{\bibfnamefont{R.~M.} \bibnamefont{Godbole}},
  \bibinfo{author}{\bibfnamefont{.}~\bibnamefont{Miller}, \bibfnamefont{D.J.}},
  \bibnamefont{and} \bibinfo{author}{\bibfnamefont{M.~M.}
  \bibnamefont{Muhlleitner}}, \bibinfo{journal}{JHEP}
  \textbf{\bibinfo{volume}{0712}}, \bibinfo{pages}{031} (\bibinfo{year}{2007}),
  \eprint{0708.0458}.

\bibitem[{\citenamefont{Kovalchuk}(2008)}]{Kovalchuk:2008zz}
\bibinfo{author}{\bibfnamefont{V.}~\bibnamefont{Kovalchuk}},
  \bibinfo{journal}{J.Exp.Theor.Phys.} \textbf{\bibinfo{volume}{107}},
  \bibinfo{pages}{774} (\bibinfo{year}{2008}).

\bibitem[{\citenamefont{Cao et~al.}(2010)\citenamefont{Cao, Jackson, Keung,
  Low, and Shu}}]{Cao:2009ah}
\bibinfo{author}{\bibfnamefont{Q.-H.} \bibnamefont{Cao}},
  \bibinfo{author}{\bibfnamefont{C.}~\bibnamefont{Jackson}},
  \bibinfo{author}{\bibfnamefont{W.-Y.} \bibnamefont{Keung}},
  \bibinfo{author}{\bibfnamefont{I.}~\bibnamefont{Low}}, \bibnamefont{and}
  \bibinfo{author}{\bibfnamefont{J.}~\bibnamefont{Shu}},
  \bibinfo{journal}{Phys.Rev.} \textbf{\bibinfo{volume}{D81}},
  \bibinfo{pages}{015010} (\bibinfo{year}{2010}), \eprint{0911.3398}.

\bibitem[{\citenamefont{Gao et~al.}(2010)\citenamefont{Gao, Gritsan, Guo,
  Melnikov, Schulze et~al.}}]{Gao:2010qx}
\bibinfo{author}{\bibfnamefont{Y.}~\bibnamefont{Gao}},
  \bibinfo{author}{\bibfnamefont{A.~V.} \bibnamefont{Gritsan}},
  \bibinfo{author}{\bibfnamefont{Z.}~\bibnamefont{Guo}},
  \bibinfo{author}{\bibfnamefont{K.}~\bibnamefont{Melnikov}},
  \bibinfo{author}{\bibfnamefont{M.}~\bibnamefont{Schulze}},
  \bibnamefont{et~al.}, \bibinfo{journal}{Phys.Rev.}
  \textbf{\bibinfo{volume}{D81}}, \bibinfo{pages}{075022}
  (\bibinfo{year}{2010}), \eprint{1001.3396}.

\bibitem[{\citenamefont{De~Rujula et~al.}(2010)\citenamefont{De~Rujula, Lykken,
  Pierini, Rogan, and Spiropulu}}]{DeRujula:2010ys}
\bibinfo{author}{\bibfnamefont{A.}~\bibnamefont{De~Rujula}},
  \bibinfo{author}{\bibfnamefont{J.}~\bibnamefont{Lykken}},
  \bibinfo{author}{\bibfnamefont{M.}~\bibnamefont{Pierini}},
  \bibinfo{author}{\bibfnamefont{C.}~\bibnamefont{Rogan}}, \bibnamefont{and}
  \bibinfo{author}{\bibfnamefont{M.}~\bibnamefont{Spiropulu}},
  \bibinfo{journal}{Phys.Rev.} \textbf{\bibinfo{volume}{D82}},
  \bibinfo{pages}{013003} (\bibinfo{year}{2010}), \eprint{1001.5300}.

\bibitem[{\citenamefont{Gainer et~al.}(2011)\citenamefont{Gainer, Kumar, Low,
  and Vega-Morales}}]{Gainer:2011xz}
\bibinfo{author}{\bibfnamefont{J.~S.} \bibnamefont{Gainer}},
  \bibinfo{author}{\bibfnamefont{K.}~\bibnamefont{Kumar}},
  \bibinfo{author}{\bibfnamefont{I.}~\bibnamefont{Low}}, \bibnamefont{and}
  \bibinfo{author}{\bibfnamefont{R.}~\bibnamefont{Vega-Morales}},
  \bibinfo{journal}{JHEP} \textbf{\bibinfo{volume}{1111}}, \bibinfo{pages}{027}
  (\bibinfo{year}{2011}), \eprint{1108.2274}.

\bibitem[{\citenamefont{Campbell
  et~al.}(2012{\natexlab{a}})\citenamefont{Campbell, Giele, and
  Williams}}]{Campbell:2012cz}
\bibinfo{author}{\bibfnamefont{J.~M.} \bibnamefont{Campbell}},
  \bibinfo{author}{\bibfnamefont{W.~T.} \bibnamefont{Giele}}, \bibnamefont{and}
  \bibinfo{author}{\bibfnamefont{C.}~\bibnamefont{Williams}}
  (\bibinfo{year}{2012}{\natexlab{a}}), \eprint{1204.4424}.

\bibitem[{\citenamefont{Campbell
  et~al.}(2012{\natexlab{b}})\citenamefont{Campbell, Giele, and
  Williams}}]{Campbell:2012ct}
\bibinfo{author}{\bibfnamefont{J.~M.} \bibnamefont{Campbell}},
  \bibinfo{author}{\bibfnamefont{W.~T.} \bibnamefont{Giele}}, \bibnamefont{and}
  \bibinfo{author}{\bibfnamefont{C.}~\bibnamefont{Williams}}
  (\bibinfo{year}{2012}{\natexlab{b}}), \eprint{1205.3434}.

\bibitem[{\citenamefont{Belyaev et~al.}(2012)\citenamefont{Belyaev,
  Christensen, and Pukhov}}]{Belyaev:2012qa}
\bibinfo{author}{\bibfnamefont{A.}~\bibnamefont{Belyaev}},
  \bibinfo{author}{\bibfnamefont{N.~D.} \bibnamefont{Christensen}},
  \bibnamefont{and} \bibinfo{author}{\bibfnamefont{A.}~\bibnamefont{Pukhov}}
  (\bibinfo{year}{2012}), \eprint{1207.6082}.

\bibitem[{\citenamefont{Coleppa et~al.}(2012)\citenamefont{Coleppa, Kumar, and
  Logan}}]{Coleppa:2012eh}
\bibinfo{author}{\bibfnamefont{B.}~\bibnamefont{Coleppa}},
  \bibinfo{author}{\bibfnamefont{K.}~\bibnamefont{Kumar}}, \bibnamefont{and}
  \bibinfo{author}{\bibfnamefont{H.~E.} \bibnamefont{Logan}}
  (\bibinfo{year}{2012}), \eprint{1208.2692}.

\bibitem[{\citenamefont{Bolognesi et~al.}(2012)\citenamefont{Bolognesi, Gao,
  Gritsan, Melnikov, Schulze et~al.}}]{Bolognesi:2012mm}
\bibinfo{author}{\bibfnamefont{S.}~\bibnamefont{Bolognesi}},
  \bibinfo{author}{\bibfnamefont{Y.}~\bibnamefont{Gao}},
  \bibinfo{author}{\bibfnamefont{A.~V.} \bibnamefont{Gritsan}},
  \bibinfo{author}{\bibfnamefont{K.}~\bibnamefont{Melnikov}},
  \bibinfo{author}{\bibfnamefont{M.}~\bibnamefont{Schulze}},
  \bibnamefont{et~al.} (\bibinfo{year}{2012}), \eprint{1208.4018}.

\bibitem[{\citenamefont{Boughezal et~al.}(2012)\citenamefont{Boughezal,
  LeCompte, and Petriello}}]{Boughezal:2012tz}
\bibinfo{author}{\bibfnamefont{R.}~\bibnamefont{Boughezal}},
  \bibinfo{author}{\bibfnamefont{T.~J.} \bibnamefont{LeCompte}},
  \bibnamefont{and} \bibinfo{author}{\bibfnamefont{F.}~\bibnamefont{Petriello}}
  (\bibinfo{year}{2012}), \eprint{1208.4311}.

\bibitem[{\citenamefont{Stolarski and Vega-Morales}(2012)}]{Stolarski:2012ps}
\bibinfo{author}{\bibfnamefont{D.}~\bibnamefont{Stolarski}} \bibnamefont{and}
  \bibinfo{author}{\bibfnamefont{R.}~\bibnamefont{Vega-Morales}},
  \bibinfo{journal}{Phys.Rev.} \textbf{\bibinfo{volume}{D86}},
  \bibinfo{pages}{117504} (\bibinfo{year}{2012}), \eprint{1208.4840}.

\bibitem[{\citenamefont{Avery et~al.}(2012)\citenamefont{Avery, Bourilkov,
  Chen, Cheng, Drozdetskiy et~al.}}]{Avery:2012um}
\bibinfo{author}{\bibfnamefont{P.}~\bibnamefont{Avery}},
  \bibinfo{author}{\bibfnamefont{D.}~\bibnamefont{Bourilkov}},
  \bibinfo{author}{\bibfnamefont{M.}~\bibnamefont{Chen}},
  \bibinfo{author}{\bibfnamefont{T.}~\bibnamefont{Cheng}},
  \bibinfo{author}{\bibfnamefont{A.}~\bibnamefont{Drozdetskiy}},
  \bibnamefont{et~al.} (\bibinfo{year}{2012}), \eprint{1210.0896}.

\bibitem[{\citenamefont{Chen et~al.}(2013{\natexlab{a}})\citenamefont{Chen,
  Tran, and Vega-Morales}}]{Chen:2012jy}
\bibinfo{author}{\bibfnamefont{Y.}~\bibnamefont{Chen}},
  \bibinfo{author}{\bibfnamefont{N.}~\bibnamefont{Tran}}, \bibnamefont{and}
  \bibinfo{author}{\bibfnamefont{R.}~\bibnamefont{Vega-Morales}},
  \bibinfo{journal}{JHEP} \textbf{\bibinfo{volume}{1301}}, \bibinfo{pages}{182}
  (\bibinfo{year}{2013}{\natexlab{a}}), \eprint{1211.1959}.

\bibitem[{\citenamefont{Modak et~al.}(2013)\citenamefont{Modak, Sahoo, Sinha,
  and Cheng}}]{Modak:2013sb}
\bibinfo{author}{\bibfnamefont{T.}~\bibnamefont{Modak}},
  \bibinfo{author}{\bibfnamefont{D.}~\bibnamefont{Sahoo}},
  \bibinfo{author}{\bibfnamefont{R.}~\bibnamefont{Sinha}}, \bibnamefont{and}
  \bibinfo{author}{\bibfnamefont{H.-Y.} \bibnamefont{Cheng}}
  (\bibinfo{year}{2013}), \eprint{1301.5404}.

\bibitem[{\citenamefont{Gainer et~al.}(2013)\citenamefont{Gainer, Lykken,
  Matchev, Mrenna, and Park}}]{Gainer:2013rxa}
\bibinfo{author}{\bibfnamefont{J.~S.} \bibnamefont{Gainer}},
  \bibinfo{author}{\bibfnamefont{J.}~\bibnamefont{Lykken}},
  \bibinfo{author}{\bibfnamefont{K.~T.} \bibnamefont{Matchev}},
  \bibinfo{author}{\bibfnamefont{S.}~\bibnamefont{Mrenna}}, \bibnamefont{and}
  \bibinfo{author}{\bibfnamefont{M.}~\bibnamefont{Park}},
  \bibinfo{journal}{Phys.Rev.Lett.} \textbf{\bibinfo{volume}{111}},
  \bibinfo{pages}{041801} (\bibinfo{year}{2013}), \eprint{1304.4936}.

\bibitem[{\citenamefont{Grinstein et~al.}(2013)\citenamefont{Grinstein, Murphy,
  and Pirtskhalava}}]{Grinstein:2013vsa}
\bibinfo{author}{\bibfnamefont{B.}~\bibnamefont{Grinstein}},
  \bibinfo{author}{\bibfnamefont{C.~W.} \bibnamefont{Murphy}},
  \bibnamefont{and}
  \bibinfo{author}{\bibfnamefont{D.}~\bibnamefont{Pirtskhalava}},
  \bibinfo{journal}{JHEP} \textbf{\bibinfo{volume}{1310}}, \bibinfo{pages}{077}
  (\bibinfo{year}{2013}), \eprint{1305.6938}.

\bibitem[{\citenamefont{Sun et~al.}(2013)\citenamefont{Sun, Wang, and
  Gao}}]{Sun:2013yra}
\bibinfo{author}{\bibfnamefont{Y.}~\bibnamefont{Sun}},
  \bibinfo{author}{\bibfnamefont{X.-F.} \bibnamefont{Wang}}, \bibnamefont{and}
  \bibinfo{author}{\bibfnamefont{D.-N.} \bibnamefont{Gao}}
  (\bibinfo{year}{2013}), \eprint{1309.4171}.

\bibitem[{\citenamefont{Anderson et~al.}(2013)\citenamefont{Anderson,
  Bolognesi, Caola, Gao, Gritsan et~al.}}]{Anderson:2013fba}
\bibinfo{author}{\bibfnamefont{I.}~\bibnamefont{Anderson}},
  \bibinfo{author}{\bibfnamefont{S.}~\bibnamefont{Bolognesi}},
  \bibinfo{author}{\bibfnamefont{F.}~\bibnamefont{Caola}},
  \bibinfo{author}{\bibfnamefont{Y.}~\bibnamefont{Gao}},
  \bibinfo{author}{\bibfnamefont{A.~V.} \bibnamefont{Gritsan}},
  \bibnamefont{et~al.} (\bibinfo{year}{2013}), \eprint{1309.4819}.

\bibitem[{\citenamefont{Chen et~al.}(2013{\natexlab{b}})\citenamefont{Chen,
  Cheng, Gainer, Korytov, Matchev et~al.}}]{Chen:2013waa}
\bibinfo{author}{\bibfnamefont{M.}~\bibnamefont{Chen}},
  \bibinfo{author}{\bibfnamefont{T.}~\bibnamefont{Cheng}},
  \bibinfo{author}{\bibfnamefont{J.~S.} \bibnamefont{Gainer}},
  \bibinfo{author}{\bibfnamefont{A.}~\bibnamefont{Korytov}},
  \bibinfo{author}{\bibfnamefont{K.~T.} \bibnamefont{Matchev}},
  \bibnamefont{et~al.} (\bibinfo{year}{2013}{\natexlab{b}}),
  \eprint{1310.1397}.

\bibitem[{\citenamefont{Buchalla et~al.}(2013)\citenamefont{Buchalla, Cata, and
  D'Ambrosio}}]{Buchalla:2013mpa}
\bibinfo{author}{\bibfnamefont{G.}~\bibnamefont{Buchalla}},
  \bibinfo{author}{\bibfnamefont{O.}~\bibnamefont{Cata}}, \bibnamefont{and}
  \bibinfo{author}{\bibfnamefont{G.}~\bibnamefont{D'Ambrosio}}
  (\bibinfo{year}{2013}), \eprint{1310.2574}.

\bibitem[{\citenamefont{Chen and Vega-Morales}(2014)}]{Chen:2013ejz}
\bibinfo{author}{\bibfnamefont{Y.}~\bibnamefont{Chen}} \bibnamefont{and}
  \bibinfo{author}{\bibfnamefont{R.}~\bibnamefont{Vega-Morales}},
  \bibinfo{journal}{JHEP} \textbf{\bibinfo{volume}{1404}}, \bibinfo{pages}{057}
  (\bibinfo{year}{2014}), \eprint{1310.2893}.

\bibitem[{\citenamefont{Gainer et~al.}(2014)\citenamefont{Gainer, Lykken,
  Matchev, Mrenna, and Park}}]{Gainer:2014hha}
\bibinfo{author}{\bibfnamefont{J.~S.} \bibnamefont{Gainer}},
  \bibinfo{author}{\bibfnamefont{J.}~\bibnamefont{Lykken}},
  \bibinfo{author}{\bibfnamefont{K.~T.} \bibnamefont{Matchev}},
  \bibinfo{author}{\bibfnamefont{S.}~\bibnamefont{Mrenna}}, \bibnamefont{and}
  \bibinfo{author}{\bibfnamefont{M.}~\bibnamefont{Park}}
  (\bibinfo{year}{2014}), \eprint{1403.4951}.

\bibitem[{\citenamefont{Chen et~al.}(2014{\natexlab{a}})\citenamefont{Chen,
  Harnik, and Vega-Morales}}]{Chen:2014gka}
\bibinfo{author}{\bibfnamefont{Y.}~\bibnamefont{Chen}},
  \bibinfo{author}{\bibfnamefont{R.}~\bibnamefont{Harnik}}, \bibnamefont{and}
  \bibinfo{author}{\bibfnamefont{R.}~\bibnamefont{Vega-Morales}},
  \bibinfo{journal}{Phys.Rev.Lett.} \textbf{\bibinfo{volume}{113}},
  \bibinfo{pages}{191801} (\bibinfo{year}{2014}{\natexlab{a}}),
  \eprint{1404.1336}.

\bibitem[{\citenamefont{Chen et~al.}(2015{\natexlab{a}})\citenamefont{Chen,
  Di~Marco, Lykken, Spiropulu, Vega-Morales et~al.}}]{Chen:2014pia}
\bibinfo{author}{\bibfnamefont{Y.}~\bibnamefont{Chen}},
  \bibinfo{author}{\bibfnamefont{E.}~\bibnamefont{Di~Marco}},
  \bibinfo{author}{\bibfnamefont{J.}~\bibnamefont{Lykken}},
  \bibinfo{author}{\bibfnamefont{M.}~\bibnamefont{Spiropulu}},
  \bibinfo{author}{\bibfnamefont{R.}~\bibnamefont{Vega-Morales}},
  \bibnamefont{et~al.}, \bibinfo{journal}{JHEP}
  \textbf{\bibinfo{volume}{1501}}, \bibinfo{pages}{125}
  (\bibinfo{year}{2015}{\natexlab{a}}), \eprint{1401.2077}.

\bibitem[{\citenamefont{Chen et~al.}(2015{\natexlab{b}})\citenamefont{Chen,
  Harnik, and Vega-Morales}}]{Chen:2015iha}
\bibinfo{author}{\bibfnamefont{Y.}~\bibnamefont{Chen}},
  \bibinfo{author}{\bibfnamefont{R.}~\bibnamefont{Harnik}}, \bibnamefont{and}
  \bibinfo{author}{\bibfnamefont{R.}~\bibnamefont{Vega-Morales}}
  (\bibinfo{year}{2015}{\natexlab{b}}), \eprint{1503.05855}.

\bibitem[{\citenamefont{Bhattacherjee et~al.}(2015)\citenamefont{Bhattacherjee,
  Modak, Patra, and Sinha}}]{Bhattacherjee:2015xra}
\bibinfo{author}{\bibfnamefont{B.}~\bibnamefont{Bhattacherjee}},
  \bibinfo{author}{\bibfnamefont{T.}~\bibnamefont{Modak}},
  \bibinfo{author}{\bibfnamefont{S.~K.} \bibnamefont{Patra}}, \bibnamefont{and}
  \bibinfo{author}{\bibfnamefont{R.}~\bibnamefont{Sinha}}
  (\bibinfo{year}{2015}), \eprint{1503.08924}.

\bibitem[{\citenamefont{Gonzalez-Alonso
  et~al.}(2015)\citenamefont{Gonzalez-Alonso, Greljo, Isidori, and
  Marzocca}}]{Gonzalez-Alonso:2015bha}
\bibinfo{author}{\bibfnamefont{M.}~\bibnamefont{Gonzalez-Alonso}},
  \bibinfo{author}{\bibfnamefont{A.}~\bibnamefont{Greljo}},
  \bibinfo{author}{\bibfnamefont{G.}~\bibnamefont{Isidori}}, \bibnamefont{and}
  \bibinfo{author}{\bibfnamefont{D.}~\bibnamefont{Marzocca}}
  (\bibinfo{year}{2015}), \eprint{1504.04018}.

\bibitem[{\citenamefont{Ginzburg et~al.}(2001)\citenamefont{Ginzburg, Krawczyk,
  and Osland}}]{Ginzburg:2001ss}
\bibinfo{author}{\bibfnamefont{I.~F.} \bibnamefont{Ginzburg}},
  \bibinfo{author}{\bibfnamefont{M.}~\bibnamefont{Krawczyk}}, \bibnamefont{and}
  \bibinfo{author}{\bibfnamefont{P.}~\bibnamefont{Osland}}
  (\bibinfo{year}{2001}), \eprint{hep-ph/0101208}.

\bibitem[{\citenamefont{Dumont et~al.}(2014)\citenamefont{Dumont, Gunion,
  Jiang, and Kraml}}]{Dumont:2014wha}
\bibinfo{author}{\bibfnamefont{B.}~\bibnamefont{Dumont}},
  \bibinfo{author}{\bibfnamefont{J.~F.} \bibnamefont{Gunion}},
  \bibinfo{author}{\bibfnamefont{Y.}~\bibnamefont{Jiang}}, \bibnamefont{and}
  \bibinfo{author}{\bibfnamefont{S.}~\bibnamefont{Kraml}},
  \bibinfo{journal}{Phys.Rev.} \textbf{\bibinfo{volume}{D90}},
  \bibinfo{pages}{035021} (\bibinfo{year}{2014}), \eprint{1405.3584}.

\bibitem[{\citenamefont{El~Hedri et~al.}(2013)\citenamefont{El~Hedri, Fox, and
  Wacker}}]{Hedri:2013wea}
\bibinfo{author}{\bibfnamefont{S.}~\bibnamefont{El~Hedri}},
  \bibinfo{author}{\bibfnamefont{P.~J.} \bibnamefont{Fox}}, \bibnamefont{and}
  \bibinfo{author}{\bibfnamefont{J.~G.} \bibnamefont{Wacker}}
  (\bibinfo{year}{2013}), \eprint{1311.6488}.

\bibitem[{\citenamefont{Chen et~al.}(2014{\natexlab{b}})\citenamefont{Chen,
  Di~Marco, Lykken, Spiropulu, Vega-Morales et~al.}}]{Chen:2014hqs}
\bibinfo{author}{\bibfnamefont{Y.}~\bibnamefont{Chen}},
  \bibinfo{author}{\bibfnamefont{E.}~\bibnamefont{Di~Marco}},
  \bibinfo{author}{\bibfnamefont{J.}~\bibnamefont{Lykken}},
  \bibinfo{author}{\bibfnamefont{M.}~\bibnamefont{Spiropulu}},
  \bibinfo{author}{\bibfnamefont{R.}~\bibnamefont{Vega-Morales}},
  \bibnamefont{et~al.} (\bibinfo{year}{2014}{\natexlab{b}}),
  \eprint{1410.4817}.

\bibitem[{\citenamefont{Schael et~al.}(2006)}]{ALEPH:2005ab}
\bibinfo{author}{\bibfnamefont{S.}~\bibnamefont{Schael}} \bibnamefont{et~al.}
  (\bibinfo{collaboration}{ALEPH, DELPHI, L3, OPAL, SLD, LEP Electroweak
  Working Group, SLD Electroweak Group, SLD Heavy Flavour Group}),
  \bibinfo{journal}{Phys.Rept.} \textbf{\bibinfo{volume}{427}},
  \bibinfo{pages}{257} (\bibinfo{year}{2006}), \eprint{hep-ex/0509008}.

\bibitem[{\citenamefont{Schael et~al.}(2013)}]{Schael:2013ita}
\bibinfo{author}{\bibfnamefont{S.}~\bibnamefont{Schael}} \bibnamefont{et~al.}
  (\bibinfo{collaboration}{ALEPH, DELPHI, L3, OPAL, LEP Electroweak}),
  \bibinfo{journal}{Phys.Rept.} \textbf{\bibinfo{volume}{532}},
  \bibinfo{pages}{119} (\bibinfo{year}{2013}), \eprint{1302.3415}.

\bibitem[{\citenamefont{Chatrchyan
  et~al.}(2014{\natexlab{b}})}]{Chatrchyan:2013fya}
\bibinfo{author}{\bibfnamefont{S.}~\bibnamefont{Chatrchyan}}
  \bibnamefont{et~al.} (\bibinfo{collaboration}{CMS}),
  \bibinfo{journal}{Phys.Rev.} \textbf{\bibinfo{volume}{D89}},
  \bibinfo{pages}{092005} (\bibinfo{year}{2014}{\natexlab{b}}),
  \eprint{1308.6832}.

\bibitem[{\citenamefont{Aad et~al.}(2015)}]{Aad:2014mda}
\bibinfo{author}{\bibfnamefont{G.}~\bibnamefont{Aad}} \bibnamefont{et~al.}
  (\bibinfo{collaboration}{ATLAS}), \bibinfo{journal}{JHEP}
  \textbf{\bibinfo{volume}{1501}}, \bibinfo{pages}{049} (\bibinfo{year}{2015}),
  \eprint{1410.7238}.

\bibitem[{\citenamefont{Bredenstein
  et~al.}(2006{\natexlab{a}})\citenamefont{Bredenstein, Denner, Dittmaier, and
  Weber}}]{Bredenstein:2006rh}
\bibinfo{author}{\bibfnamefont{A.}~\bibnamefont{Bredenstein}},
  \bibinfo{author}{\bibfnamefont{A.}~\bibnamefont{Denner}},
  \bibinfo{author}{\bibfnamefont{S.}~\bibnamefont{Dittmaier}},
  \bibnamefont{and} \bibinfo{author}{\bibfnamefont{M.}~\bibnamefont{Weber}},
  \bibinfo{journal}{Phys.Rev.} \textbf{\bibinfo{volume}{D74}},
  \bibinfo{pages}{013004} (\bibinfo{year}{2006}{\natexlab{a}}),
  \eprint{hep-ph/0604011}.

\bibitem[{\citenamefont{Bredenstein
  et~al.}(2006{\natexlab{b}})\citenamefont{Bredenstein, Denner, Dittmaier, and
  Weber}}]{Bredenstein:2006nk}
\bibinfo{author}{\bibfnamefont{A.}~\bibnamefont{Bredenstein}},
  \bibinfo{author}{\bibfnamefont{A.}~\bibnamefont{Denner}},
  \bibinfo{author}{\bibfnamefont{S.}~\bibnamefont{Dittmaier}},
  \bibnamefont{and} \bibinfo{author}{\bibfnamefont{M.}~\bibnamefont{Weber}},
  \bibinfo{journal}{Nucl.Phys.Proc.Suppl.} \textbf{\bibinfo{volume}{160}},
  \bibinfo{pages}{131} (\bibinfo{year}{2006}{\natexlab{b}}),
  \eprint{hep-ph/0607060}.

\bibitem[{\citenamefont{Boselli et~al.}(2015)\citenamefont{Boselli, Calame,
  Montagna, Nicrosini, and Piccinini}}]{Boselli:2015aha}
\bibinfo{author}{\bibfnamefont{S.}~\bibnamefont{Boselli}},
  \bibinfo{author}{\bibfnamefont{C.~M.~C.} \bibnamefont{Calame}},
  \bibinfo{author}{\bibfnamefont{G.}~\bibnamefont{Montagna}},
  \bibinfo{author}{\bibfnamefont{O.}~\bibnamefont{Nicrosini}},
  \bibnamefont{and} \bibinfo{author}{\bibfnamefont{F.}~\bibnamefont{Piccinini}}
  (\bibinfo{year}{2015}), \eprint{1503.07394}.

\bibitem[{\citenamefont{Gonzalez-Alonso
  et~al.}(2014)\citenamefont{Gonzalez-Alonso, Greljo, Isidori, and
  Marzocca}}]{Gonzalez-Alonso:2014eva}
\bibinfo{author}{\bibfnamefont{M.}~\bibnamefont{Gonzalez-Alonso}},
  \bibinfo{author}{\bibfnamefont{A.}~\bibnamefont{Greljo}},
  \bibinfo{author}{\bibfnamefont{G.}~\bibnamefont{Isidori}}, \bibnamefont{and}
  \bibinfo{author}{\bibfnamefont{D.}~\bibnamefont{Marzocca}}
  (\bibinfo{year}{2014}), \eprint{1412.6038}.

\bibitem[{\citenamefont{Khachatryan
  et~al.}(2014{\natexlab{c}})}]{Khachatryan:2014kca}
\bibinfo{author}{\bibfnamefont{V.}~\bibnamefont{Khachatryan}}
  \bibnamefont{et~al.} (\bibinfo{collaboration}{CMS Collaboration})
  (\bibinfo{year}{2014}{\natexlab{c}}), \eprint{1411.3441}.

\bibitem[{\citenamefont{Chen et~al.}(2015{\natexlab{c}})\citenamefont{Chen,
  Stolarski, Vega-Morales et~al.}}]{followup2}
\bibinfo{author}{\bibfnamefont{Y.}~\bibnamefont{Chen}},
  \bibinfo{author}{\bibfnamefont{D.}~\bibnamefont{Stolarski}},
  \bibinfo{author}{\bibfnamefont{R.}~\bibnamefont{Vega-Morales}},
  \bibnamefont{et~al.} (\bibinfo{year}{2015}{\natexlab{c}}), \eprint{Work in
  progress}.

\bibitem[{\citenamefont{Cahn et~al.}(1979)\citenamefont{Cahn, Chanowitz, and
  Fleishon}}]{Cahn:1978nz}
\bibinfo{author}{\bibfnamefont{R.}~\bibnamefont{Cahn}},
  \bibinfo{author}{\bibfnamefont{M.~S.} \bibnamefont{Chanowitz}},
  \bibnamefont{and} \bibinfo{author}{\bibfnamefont{N.}~\bibnamefont{Fleishon}},
  \bibinfo{journal}{Phys.Lett.} \textbf{\bibinfo{volume}{B82}},
  \bibinfo{pages}{113} (\bibinfo{year}{1979}).

\bibitem[{\citenamefont{Bergstrom and Hulth}(1985)}]{Bergstrom:1985hp}
\bibinfo{author}{\bibfnamefont{L.}~\bibnamefont{Bergstrom}} \bibnamefont{and}
  \bibinfo{author}{\bibfnamefont{G.}~\bibnamefont{Hulth}},
  \bibinfo{journal}{Nucl.Phys.} \textbf{\bibinfo{volume}{B259}},
  \bibinfo{pages}{137} (\bibinfo{year}{1985}).

\bibitem[{\citenamefont{Ellis et~al.}(1976)\citenamefont{Ellis, Gaillard, and
  Nanopoulos}}]{Ellis:1975ap}
\bibinfo{author}{\bibfnamefont{J.~R.} \bibnamefont{Ellis}},
  \bibinfo{author}{\bibfnamefont{M.~K.} \bibnamefont{Gaillard}},
  \bibnamefont{and} \bibinfo{author}{\bibfnamefont{D.~V.}
  \bibnamefont{Nanopoulos}}, \bibinfo{journal}{Nucl.Phys.}
  \textbf{\bibinfo{volume}{B106}}, \bibinfo{pages}{292} (\bibinfo{year}{1976}).

\bibitem[{\citenamefont{Shifman et~al.}(1979)\citenamefont{Shifman, Vainshtein,
  Voloshin, and Zakharov}}]{Shifman:1979eb}
\bibinfo{author}{\bibfnamefont{M.~A.} \bibnamefont{Shifman}},
  \bibinfo{author}{\bibfnamefont{A.}~\bibnamefont{Vainshtein}},
  \bibinfo{author}{\bibfnamefont{M.}~\bibnamefont{Voloshin}}, \bibnamefont{and}
  \bibinfo{author}{\bibfnamefont{V.~I.} \bibnamefont{Zakharov}},
  \bibinfo{journal}{Sov.J.Nucl.Phys.} \textbf{\bibinfo{volume}{30}},
  \bibinfo{pages}{711} (\bibinfo{year}{1979}).

\bibitem[{\citenamefont{Weiler and Yuan}(1989)}]{Weiler:1988xn}
\bibinfo{author}{\bibfnamefont{T.~J.} \bibnamefont{Weiler}} \bibnamefont{and}
  \bibinfo{author}{\bibfnamefont{T.-C.} \bibnamefont{Yuan}},
  \bibinfo{journal}{Nucl.Phys.} \textbf{\bibinfo{volume}{B318}},
  \bibinfo{pages}{337} (\bibinfo{year}{1989}).

\bibitem[{\citenamefont{Djouadi}(2008{\natexlab{a}})}]{Djouadi:2005gi}
\bibinfo{author}{\bibfnamefont{A.}~\bibnamefont{Djouadi}},
  \bibinfo{journal}{Phys.Rept.} \textbf{\bibinfo{volume}{457}},
  \bibinfo{pages}{1} (\bibinfo{year}{2008}{\natexlab{a}}),
  \eprint{hep-ph/0503172}.

\bibitem[{\citenamefont{Djouadi}(2008{\natexlab{b}})}]{Djouadi:2005gj}
\bibinfo{author}{\bibfnamefont{A.}~\bibnamefont{Djouadi}},
  \bibinfo{journal}{Phys.Rept.} \textbf{\bibinfo{volume}{459}},
  \bibinfo{pages}{1} (\bibinfo{year}{2008}{\natexlab{b}}),
  \eprint{hep-ph/0503173}.

\bibitem[{\citenamefont{Chen et~al.}(2014{\natexlab{c}})\citenamefont{Chen,
  Falkowski, Low, and Vega-Morales}}]{Chen:2014ona}
\bibinfo{author}{\bibfnamefont{Y.}~\bibnamefont{Chen}},
  \bibinfo{author}{\bibfnamefont{A.}~\bibnamefont{Falkowski}},
  \bibinfo{author}{\bibfnamefont{I.}~\bibnamefont{Low}}, \bibnamefont{and}
  \bibinfo{author}{\bibfnamefont{R.}~\bibnamefont{Vega-Morales}},
  \bibinfo{journal}{Phys.Rev.} \textbf{\bibinfo{volume}{D90}},
  \bibinfo{pages}{113006} (\bibinfo{year}{2014}{\natexlab{c}}),
  \eprint{1405.6723}.

\bibitem[{\citenamefont{Shen and Zhu}(2015)}]{Shen:2015pha}
\bibinfo{author}{\bibfnamefont{C.}~\bibnamefont{Shen}} \bibnamefont{and}
  \bibinfo{author}{\bibfnamefont{S.-h.} \bibnamefont{Zhu}}
  (\bibinfo{year}{2015}), \eprint{1504.05626}.

\bibitem[{\citenamefont{Farina et~al.}(2015)\citenamefont{Farina, Grossman, and
  Robinson}}]{Farina:2015dua}
\bibinfo{author}{\bibfnamefont{M.}~\bibnamefont{Farina}},
  \bibinfo{author}{\bibfnamefont{Y.}~\bibnamefont{Grossman}}, \bibnamefont{and}
  \bibinfo{author}{\bibfnamefont{D.~J.} \bibnamefont{Robinson}}
  (\bibinfo{year}{2015}), \eprint{1503.06470}.

\bibitem[{\citenamefont{Lai et~al.}(2000)}]{Lai:1999wy}
\bibinfo{author}{\bibfnamefont{H.}~\bibnamefont{Lai}} \bibnamefont{et~al.}
  (\bibinfo{collaboration}{CTEQ Collaboration}), \bibinfo{journal}{Eur.Phys.J.}
  \textbf{\bibinfo{volume}{C12}}, \bibinfo{pages}{375} (\bibinfo{year}{2000}),
  \eprint{hep-ph/9903282}.

\bibitem[{\citenamefont{Pumplin et~al.}(2002)\citenamefont{Pumplin, Stump,
  Huston, Lai, Nadolsky et~al.}}]{Pumplin:2002vw}
\bibinfo{author}{\bibfnamefont{J.}~\bibnamefont{Pumplin}},
  \bibinfo{author}{\bibfnamefont{D.}~\bibnamefont{Stump}},
  \bibinfo{author}{\bibfnamefont{J.}~\bibnamefont{Huston}},
  \bibinfo{author}{\bibfnamefont{H.}~\bibnamefont{Lai}},
  \bibinfo{author}{\bibfnamefont{P.~M.} \bibnamefont{Nadolsky}},
  \bibnamefont{et~al.}, \bibinfo{journal}{JHEP}
  \textbf{\bibinfo{volume}{0207}}, \bibinfo{pages}{012} (\bibinfo{year}{2002}),
  \eprint{hep-ph/0201195}.

\bibitem[{\citenamefont{Alwall et~al.}(2014)\citenamefont{Alwall, Frederix,
  Frixione, Hirschi, Maltoni et~al.}}]{Alwall:2014hca}
\bibinfo{author}{\bibfnamefont{J.}~\bibnamefont{Alwall}},
  \bibinfo{author}{\bibfnamefont{R.}~\bibnamefont{Frederix}},
  \bibinfo{author}{\bibfnamefont{S.}~\bibnamefont{Frixione}},
  \bibinfo{author}{\bibfnamefont{V.}~\bibnamefont{Hirschi}},
  \bibinfo{author}{\bibfnamefont{F.}~\bibnamefont{Maltoni}},
  \bibnamefont{et~al.}, \bibinfo{journal}{JHEP}
  \textbf{\bibinfo{volume}{1407}}, \bibinfo{pages}{079} (\bibinfo{year}{2014}),
  \eprint{1405.0301}.

\bibitem[{CMS(2014)}]{CMS-PAS-HIG-14-014}
\bibinfo{type}{Tech. Rep.} \bibinfo{number}{CMS-PAS-HIG-14-014},
  \bibinfo{institution}{CERN}, \bibinfo{address}{Geneva}
  (\bibinfo{year}{2014}).

\bibitem[{\citenamefont{Dittmaier et~al.}(2011)}]{Dittmaier:2011ti}
\bibinfo{author}{\bibfnamefont{S.}~\bibnamefont{Dittmaier}}
  \bibnamefont{et~al.} (\bibinfo{collaboration}{LHC Higgs Cross Section Working
  Group}) (\bibinfo{year}{2011}), \eprint{1101.0593}.

\bibitem[{\citenamefont{Heinemeyer et~al.}(2013)}]{Heinemeyer:2013tqa}
\bibinfo{author}{\bibfnamefont{S.}~\bibnamefont{Heinemeyer}}
  \bibnamefont{et~al.} (\bibinfo{collaboration}{LHC Higgs Cross Section Working
  Group}) (\bibinfo{year}{2013}), \eprint{1307.1347}.

\bibitem[{\citenamefont{Khachatryan
  et~al.}(2014{\natexlab{d}})}]{Khachatryan:2014ira}
\bibinfo{author}{\bibfnamefont{V.}~\bibnamefont{Khachatryan}}
  \bibnamefont{et~al.} (\bibinfo{collaboration}{CMS}),
  \bibinfo{journal}{Eur.Phys.J.} \textbf{\bibinfo{volume}{C74}},
  \bibinfo{pages}{3076} (\bibinfo{year}{2014}{\natexlab{d}}),
  \eprint{1407.0558}.

\bibitem[{\citenamefont{Dawson et~al.}(2013)\citenamefont{Dawson, Gritsan,
  Logan, Qian, Tully et~al.}}]{Dawson:2013bba}
\bibinfo{author}{\bibfnamefont{S.}~\bibnamefont{Dawson}},
  \bibinfo{author}{\bibfnamefont{A.}~\bibnamefont{Gritsan}},
  \bibinfo{author}{\bibfnamefont{H.}~\bibnamefont{Logan}},
  \bibinfo{author}{\bibfnamefont{J.}~\bibnamefont{Qian}},
  \bibinfo{author}{\bibfnamefont{C.}~\bibnamefont{Tully}}, \bibnamefont{et~al.}
  (\bibinfo{year}{2013}), \eprint{1310.8361}.

\bibitem[{\citenamefont{ATLAS-collaboration}(2012)}]{ATLAS-collaboration:1484890}
\bibinfo{author}{\bibfnamefont{T.}~\bibnamefont{ATLAS-collaboration}},
  \bibinfo{type}{Tech. Rep.} \bibinfo{number}{ATL-PHYS-PUB-2012-004},
  \bibinfo{institution}{CERN}, \bibinfo{address}{Geneva}
  (\bibinfo{year}{2012}), \urlprefix\url{http://cds.cern.ch/record/1484890}.

\bibitem[{\citenamefont{Murray}(2013)}]{htoAA}
\bibinfo{author}{\bibfnamefont{W.}~\bibnamefont{Murray}}
  (\bibinfo{year}{2013}),
  \eprint{{http://indico.cern.ch/event/252045\\/session/3/contribution/8/material/slides/0.pdf}}.

\bibitem[{\citenamefont{Chen et~al.}(2013{\natexlab{c}})}]{WEBSITE}
\bibinfo{author}{\bibfnamefont{Y.}~\bibnamefont{Chen}} \bibnamefont{et~al.}
  (\bibinfo{year}{2013}{\natexlab{c}}), \eprint{{Website under
  construction:\\http://yichen.me/project/GoldenChannel/}}.

\end{thebibliography}

\end{document}